\documentclass{aa}

\usepackage{graphicx}
\usepackage{natbib}
\usepackage{amssymb}
\usepackage{amsmath}
\usepackage{txfonts}
\usepackage[figuresright]{rotating}
\usepackage{ulem}
\bibpunct{(}{)}{;}{a}{}{,} 

\newcommand{\ud}{\mathrm{d}} 
\newcommand{\htwo}{\mbox{${\rm H}_{2}$}}

\newcommand{\halpha}{\mbox{H$\alpha$}}
\newcommand{\palpha}{\mbox{Pa$\alpha$}}

\newcommand{\msun}{\mbox{$M_{\odot}$}}

\newcommand{\zsun}{\mbox{$Z_{\odot}$}}

\newcommand{\ebv}{\mbox{$E(B\!-\!V)$}}

\newcommand{\hst}{\emph{HST}}
\newcommand{\acs}{\emph{ACS}}
\newcommand{\wfpc}{\emph{WFPC2}}

\newcommand{\ishape}{{\sc Ishape}}

\newcommand{\iraf}{{\sc IRAF}}
\newcommand{\daophot}{{\sc DAOPHOT}}
\newcommand{\sextractor}{{\sc SExtractor}}

\newcommand{\galev}{{\sc galev}}
\newcommand{\analysed}{{\sc analysed}}

\newcommand{\kms}{\ensuremath{\mathrm{km}\,\mathrm{s}^{-1}}}
\newcommand{\uband}{\mbox{$U_{F336W}$}}
\newcommand{\bband}{\mbox{$B_{F435W}$}}
\newcommand{\vband}{\mbox{$V_{F555W}$}}
\newcommand{\iband}{\mbox{$I_{F814W}$}}

\newcommand{\tkin}{\mbox{$T_{\mathrm{kin}}$}}

\begin{document}

\title{The spatial distribution of star and cluster formation in M~51}

\titlerunning{The spatial distribution of star and cluster formation in M~51}

\author{R. A. Scheepmaker 
	\and
	H. J. G. L. M. Lamers
       \and
       P. Anders
       \and
       S. S. Larsen 
       }

\authorrunning{R. A. Scheepmaker et al.}

\offprints{R. A. Scheepmaker, \email{R.A.Scheepmaker@uu.nl}}

\institute{Astronomical Institute, Utrecht University,
  Princetonplein 5, NL-3584 CC Utrecht, The Netherlands}

\date{Received 2 October 2008 / Accepted 2 December 2008}


\abstract
{}
{We study the connection between spatially resolved star formation
  and young star clusters across the disc of M~51. }
{We combine star cluster data based on $B$, $V$, and $I$-band
  \emph{Hubble Space Telescope} \acs\ imaging, together with new
  \wfpc\ $U$-band photometry to derive ages, masses, and extinctions of 1580 resolved star
  clusters using SSP models. This data is combined with data on the spatially resolved star formation rates and gas surface densities, as well as H$\alpha$ and 20~cm radio-continuum (RC) emission, which allows us to study the spatial correlations between star formation and star clusters. Two-point autocorrelation functions are
  used to study the clustering of star clusters as a function of
  spatial scale and age.}
{We find that the clustering of star clusters among themselves
  decreases both with spatial scale and age, consistent with
  hierarchical star formation. The slope of the autocorrelation functions are consistent with projected fractal dimensions in the range of 1.2--1.6, which is similar to other galaxies, therefore suggesting that the fractal dimension of hierarchical star formation is universal. Both star and cluster formation peak at
  a galactocentric radius of $\sim$2.5 and $\sim$5~kpc, which we
  tentatively attribute to the presence of the 4:1 resonance and the
  co-rotation radius. The positions of the youngest ($<10$~Myr) star clusters show the strongest correlation with the spiral arms, H$\alpha$, and the RC emission, and these correlations decrease with age. The azimuthal distribution of clusters in terms of kinematic age away from the spiral arms indicates that the majority of the clusters formed $\sim5$--20 Myr before their parental gas cloud reached the centre of the spiral arm.}
{}


\maketitle

\section{Introduction}
  \label{sec:Introduction}

The increase in the amount of multi-wavelength data during the past few decades has been of
tremendous importance for the field of star formation. On the largest scales,
these multi-wavelength studies focussed primarily on the tight
correlation between global radio-continuum (RC) and far-infrared (FIR)
emission of galaxies (e.g. \citealt{condon92, niklas97a, niklas97b,
  yun01}) and on the global Schmidt-law, which dictates that there is
a powerlaw relation between the area-normalised star formation rate (SFR) and
total gas surface density with an index of $1.4\pm0.15$
\citep{kennicutt98b}. These correlations have been proven useful in
the derivation of global star formation rates from radio-continuum
observations (e.g. \citealt{condon92}), as well as in models of galaxy
evolution \citep{mihos94}.

More recently, these multi-wavelength correlations are exploited to
study star formation \emph{locally} across single galaxies.
This is done by either utilizing the RC-FIR correlation \citep{murphy06,
  schuster07, murphy08}, or by combining \emph{Spitzer Space
  Telescope} FIR observations with H$\alpha$ and Pa$\alpha$ data,
which allows for a more thorough extinction correction on the derived
star formation rates \citep{calzetti05, kennicutt07}.
Combined with new CO and \ion{H}{I} data, recent studies show that the
Schmidt-law is also valid \emph{locally} across the spiral galaxy M~51
\citep{schuster07, kennicutt07}. 

There is ample evidence that star formation leads to formation of star clusters (\citealt{lada03}, and references
therein). Therefore, star formation across spiral galaxies has been
studied extensively through their star cluster populations, either by
using ground-based data \citep{larsen99b, larsen00a}, or by exploiting the higher
resolution \emph{Hubble Space Telescope} (\hst) observations
(e.g. \citealt{holtzman92, meurer95, whitmore99b, degrijs03d, larsen04b}). Most of these studies, however, focus on
the star cluster properties themselves, on the connection between the
galaxy's star cluster population and SFR on \emph{global} scales, or
on their spatial correlation \citep{zhang01}.
There is still a lack of
evidence, however, on the actual \emph{fraction} of current star
formation taking place in young star clusters (although
\citealt{degrijs03d} find a lower limit of $\approx 35\%$ for the Mice
galaxies and \citealt{fall05} find a lower limit of 20\% in the
Antennae). Although it is probably true that most star formation takes
place in some sort of embedded, clustered form \citep{lada03}, it is
not clear yet which fraction of these clusters are actually observable
as star clusters after they emerge from their natal gas cloud
\citep{bastian06b, goodwin06} and if, or how, this depends on
environment. The answers to such questions could give us important
clues on the process of star formation and the early evolution of
young star clusters.

Before such a quantitative comparison can be made, however, we first need to determine the amount of \emph{spatial} correlation between star and cluster formation, since any quantitative comparison is only meaningful for regions that are spatially correlated. Besides, determining the spatial correlation between clusters themselves allows us to study the amount and evolution of clustering of star clusters, which provide important clues as to which extent the formation of clusters is hierarchical \citep[e.g.][]{efremov98, bastian07}.

For M~51 data with a wide wavelength range and good
spatial resolution are available. This makes it possible to study, first of all, how well the positions of star clusters are tracing the regions of star formation across this spiral galaxy, and second of all, which fraction of the star formation is taking place in the form of visually observable star clusters. 
In this study we combine data on the
\emph{resolved} Schmidt-law in M~51 (SFR and gas surface densities,
\citealt{kennicutt07}) as well as archival radio-continuum, 2MASS and \hst/\wfpc\ data with 
slightly resolved young star
clusters from \citet{scheepmaker07}. These datasets are covering a
large fraction of the disc, allowing us to compare star and
cluster formation  under a range of conditions, such a low/high gas and stellar surface densities, within a single galaxy consistently. 

In this work we will introduce the new datasets used, and present new \wfpc\ \uband\ photometry. This photometry allows us to add ages, masses and extinctions to the star cluster data of \citet{scheepmaker07}. In the remainder of this work we will then focus on the \emph{spatial} correlations between star and cluster formation, for differently aged cluster populations. A \emph{quantitative} comparison between star and cluster formation, as well as the cluster initial mass function (CIMF), will be the subject of a follow-up paper \citep{scheepmaker09b}.

In Sect.~\ref{sec:Observations and photometry} we describe the star
cluster data we used, as well as the \hst/\wfpc\ observations.
In Sect.~\ref{sec:Selection of the cluster samples} we estimate ages, masses and extinctions of the star clusters in order to select our cluster samples.  We
describe the data on SFRs and gas densities in Sect.~\ref{sec:Star
  formation rates, gas densities and spiral arms}, together with our
method of tracing the spiral arms. Finally, in Sect.~\ref{sec:Spatial
  distribution of star clusters and star formation} we study the
spatial distribution of star clusters and star formation in M~51 and we summarize our
main results in Sect.~\ref{sec:Summary and conclusions}.

\section{Observations and photometry}
  \label{sec:Observations and photometry}

For our star cluster data, we make use of the dataset of Scheepmaker
et al.\ (2007, from here on referred to as ``S07''), complemented with
\hst/\wfpc\ U-band (\uband) observations.

\subsection{HST/ACS observations}

 S07 used the \hst/\acs\ mosaic image of M~51 in \bband\ ($\sim$$B$),
 \vband\ ($\sim$$V$) and \iband\ ($\sim$$I$) to select star clusters
 across the complete disc of M~51, based on size measurements. For a
 complete description we refer to S07 for the star cluster data and to
 \citet{mutchler05} and the M~51 mosaic website
 (\texttt{http://archive.stsci.edu/prepds/m51/}) for the image
 processing. Below, the source selection, photometry and size
 measurements of the clusters are summarized. 

The \sextractor\ package \citep{bertin96} was used to select sources
above a local threshold, cross-correlating between the three filters.
Aperture photometry in the VEGAmag system was performed on the sources
using the \iraf\footnote{The Image Redcution and Analysis Facility
(\iraf) is distributed by the National Optical Astronomy
Observatories, which are operated by the Association of Universities
for Research in Astronomy, Inc., under cooperative agreement with the
National Science Foundation.}/\daophot\ package, using an aperture
radius of 5 pixels and a background annulus with an inner radius of 10
pixels and a width of 3 pixels. A fixed aperture correction for a 3~pc source was applied, which was respectively -0.17 
mag for \bband\ and \vband\ and -0.19 mag for \iband, and the photometry was corrected for Galactic
foreground extinction in the direction of M~51 according to Appendix B
of \citet{schlegel98}.  The star clusters, being slightly extended,
were distinguished from stars based on size measurements using the
\ishape\ routine \citep{larsen99, larsen04b}, assuming EFF~15
models \citep{elson87} for the surface brightnesss profiles of the clusters.

\subsection{HST/WFPC2 observations}

For a reliable estimation of star cluster parameters using
\emph{Simple Stellar Population} models (see Sect.~\ref{subsec:SSP model
  fitting}), \citet{anders04b} have shown that at least four passbands
are necessary, with the $U$ and $B$ bands having the highest
significance for young clusters. We therefore complemented the
\hst/\acs\ data with archival \hst/\wfpc\ observations in the
\uband\ passband ($\sim U$).

\begin{figure}
\centering \includegraphics[width=85mm]{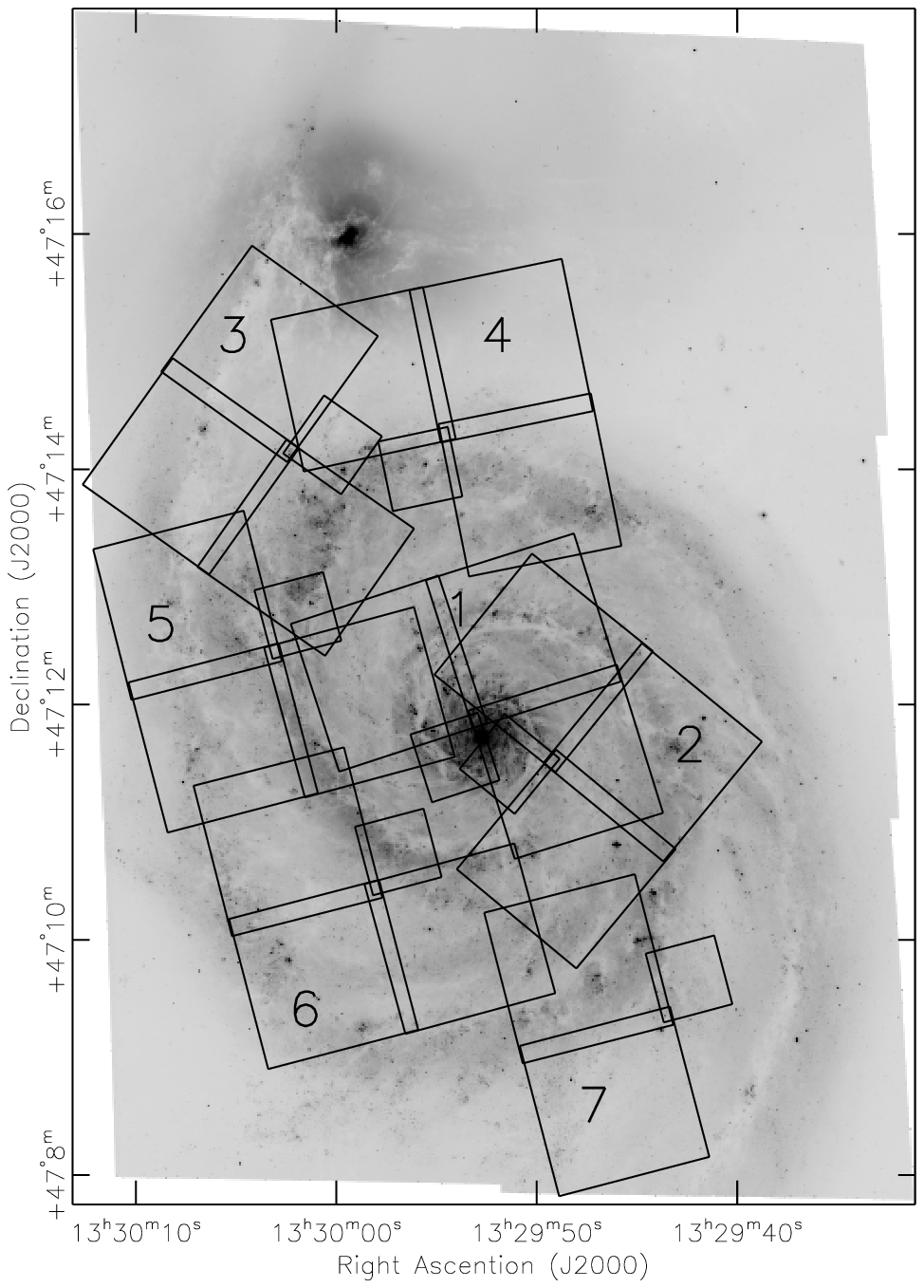}
\caption{Footprints of the \wfpc\ \uband\ fields used in this
  study, overlaid on the \acs\ \bband\ mosaic image. The numbers refer
  to the different datasets in Table~\ref{tab:observations}.}
\label{fig:footprints}
\end{figure}

A total of seven pointings were taken from the ESO/ST-ECF science
archive. The footprints of the pointings, overlaid on the
\acs\ \bband\ image, are shown in Fig.~\ref{fig:footprints}. An
overview of the different datasets is given in
Table~\ref{tab:observations}. We note that proposal ID~10501 includes
another \uband\ pointing covering the companion galaxy of M~51, NGC~5195. 
Since we are interested 
in the star clusters in the disc of M~51,
this pointing (dataset U9GA0101B) was not used.  We also did not use
the WF4 chip of dataset U9GA0601B (field 7 in
Fig~\ref{fig:footprints}), since it contained too few sources. This made the manual selection of enough reference objects impossible,
which were necessary to transform the cluster coordinates from the
\acs\ frame to the \wfpc\ frames (see below).

\begin{table*}
\caption[]{Overview of the different \hst/\wfpc\ \uband\ band datasets
  used in this study.}
\label{tab:observations}
\begin{center}
\begin{tabular}{cccccc}
\hline\hline
Field$^{a}$ & ID & PI & Dataset (association) & Exposure time & Observation date \\
\hline 
1 & 5652 & R.~P.~Kirshner & U2EW0205B & $3 \times 400 = 1200$ s & 12 May 1994 \\
\hline
2 & 7375 & N.~Scoville & U50Q0105B & $2 \times 600 = 1200$ s & 21 Jul 1999 \\
\hline
3 &  & & U9GA0201B & $2 \times 1300 = 2600$ s & 24 Jan 2006 \\
4 &  & & U9GA5301B & $2 \times 1200 = 2400$ s & 28 May 2006 \\
5 & 10501 & R.~Chandar & U9GA0401B & $2 \times 1300 = 2600$ s & 13 Nov 2005 \\
6 &  & & U9GA0501B & $2 \times 1300 = 2600$ s & 13 Nov 2005 \\
7 &  & & U9GA0601B & $2 \times 1300 = 2600$ s & 13 Nov 2005 \\
\hline
 \multicolumn{6}{l}{$^{a}$ The numbers refer to the footprints in Fig.~\ref{fig:footprints}.}
\end{tabular}
\end{center}
\end{table*}

We used the individual chips (WF and PC) of the standard
pipeline-reduced associations, i.e. no further data reduction was
performed. On every chip between 6--11 sources were matched with
sources on the ACS \bband\ image. Using these sources the coordinate
transformations between the \wfpc\ chips and the \acs\ frame were
calculated using the \iraf\ task {\sc geomap}. The task {\sc
  geoxytran} was used to transform the coordinates of the 7698
resolved sources of S07 on the \acs\ \bband\ image to the
\wfpc\ chips. In most cases the root-mean-square accuracy in X and Y
was better than one pixel, and in all cases better than 2 pixels.

Aperture photometry was performed at the transformed coordinates using
\daophot, allowing for a maximum centring shift of three pixels. For
the PC chips a $(5,10,3)$ configuration was used for the radius of
respectively the aperture, annulus and width of the annulus in
pixels. For the WF chips the configuration was $(3,7,3)$. Aperture
corrections to an aperture of 0.5\arcsec were measured using analytic
cluster profiles with an effective radius of 3~pc, which were subsequently convolved with the PSF. These corrections
for the WF and PC chips were $-0.14$ and $-0.19$ mag for the WF and PC
chips, respectively. Zeropoints (calibrated for the 0.5\arcsec aperture) and CTE
loss corrections were calculated following
\citet{dolphin00b}.\footnote{The most recent equations for the CTE
losses were taken from
\texttt{http://purcell.as.arizona.edu/wfpc2\_calib/}.}  We corrected
for Galactic foreground extinction in the direction of M~51 according
to Appendix~B of \citet{schlegel98}.

From the initial source list of 7698 resolved sources on the
\acs\ \bband\ image, we retrieved positive photometry on the
\wfpc\ \uband\ images of 5502 sources. The remaining sources had
either an aperture covering the edge of an image, a failure
in {\sc daophot's} centring algorithm or a negative flux, caused by a
combination of a faint source and noise in the background annulus. As
can be seen in Fig.~\ref{fig:footprints}, there is some overlap
between different \wfpc\ chips. If a source had photometry in
more than one \uband\ chip, the photometry with the smallest error was
selected.

\begin{figure}
\centering
 \includegraphics[width=85mm]{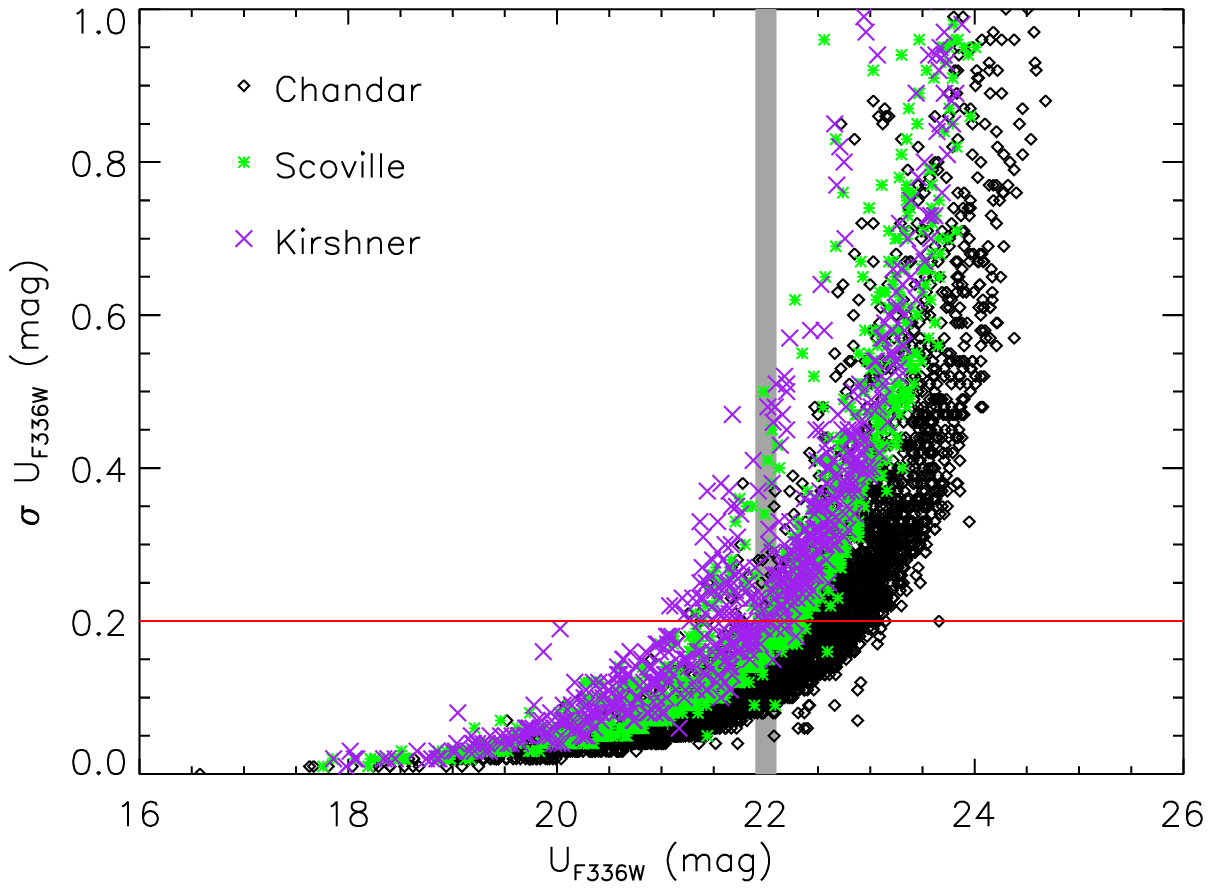} 
 \caption{The error in the \uband\ magnitude versus the
   \uband\ magnitude for all 5502 sources for which we have
   photometry. The different symbols/colours refer to the different datasets
   listed in Table~\ref{tab:observations}. The horizontal line
   indicates $\sigma \uband = 0.2$~mag and the bin at
   $\uband = 22$~mag indicates the magnitude range in which $\sim$90\% of
 the clusters has $\sigma \uband <0.2$~mag.}
\label{fig:dU_vs_U}
\end{figure}

In Fig.~\ref{fig:dU_vs_U} we show the resulting photometry and error
in the photometry for the \uband\ passband. As expected, the
photometry of fields 1 and 2 has larger errors compared to the
photometry of fields 3--7, due to the deeper exposures of the latter
fields.  The horizontal line at $\sigma(\uband) = 0.2$~mag indicates
the limit we applied to select only clusters with accurate photometry.
The bin centred on $\uband = 22$~mag indicates the
magnitude at which approximately 10\% of the clusters will be rejected
from the sample by this $\sigma(\uband) < 0.2$~mag criterion. These
limits will be used in Sect.~\ref{subsec:Selection of the different age
  samples} to select cluster samples.

\section{Selection of the cluster samples}
  \label{sec:Selection of the cluster samples}

In this study we are primarily interested in the spatial distribution
of \emph{young} star clusters in the disc of M~51. We therefore used
\emph{Simple Stellar Population} (SSP) models to estimate the age, mass and extinction of all 5502 
clusters
for which we have $UBVI$ photometry (Sect.~\ref{subsec:SSP model
  fitting}). The estimated ages were then used to select cluster
samples with different ages (Sect.~\ref{subsec:Selection of the different age
  samples}).

\subsection{SSP model fitting}
  \label{subsec:SSP model fitting}

For our age determinations we used the \galev\ SSP models
\citep{schulz02, anders03} together with the \analysed\ tool
\citep{anders04b}. In summary, \analysed\ compares the broad-band spectral energy distribution (SED, in our case UBVI) of every cluster to a large grid of modelled SEDs for different ages, extinctions and metallicities. The model with the lowest $\chi^{2}$ is selected, and the (initial) mass of the cluster is derived by scaling the entire modelled SED to the observed SED. The models and the tool have been extensively tested \citep[e.g.][]{anders04b, degrijs05, degrijs06} and the uncertainty in the derived ages is typically $<0.4$~dex. 

\begin{figure*}
\centering
 \includegraphics[width=85mm]{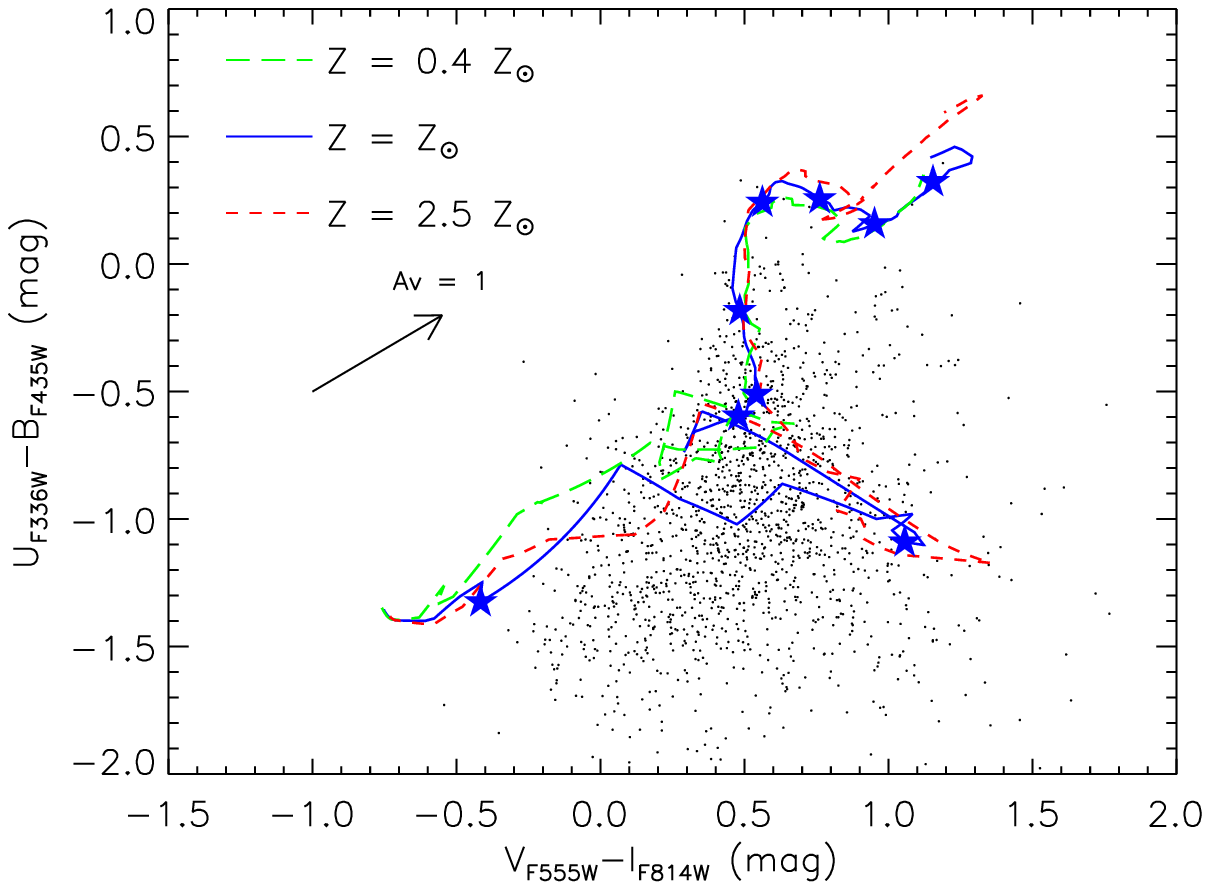} 
 \includegraphics[width=85mm]{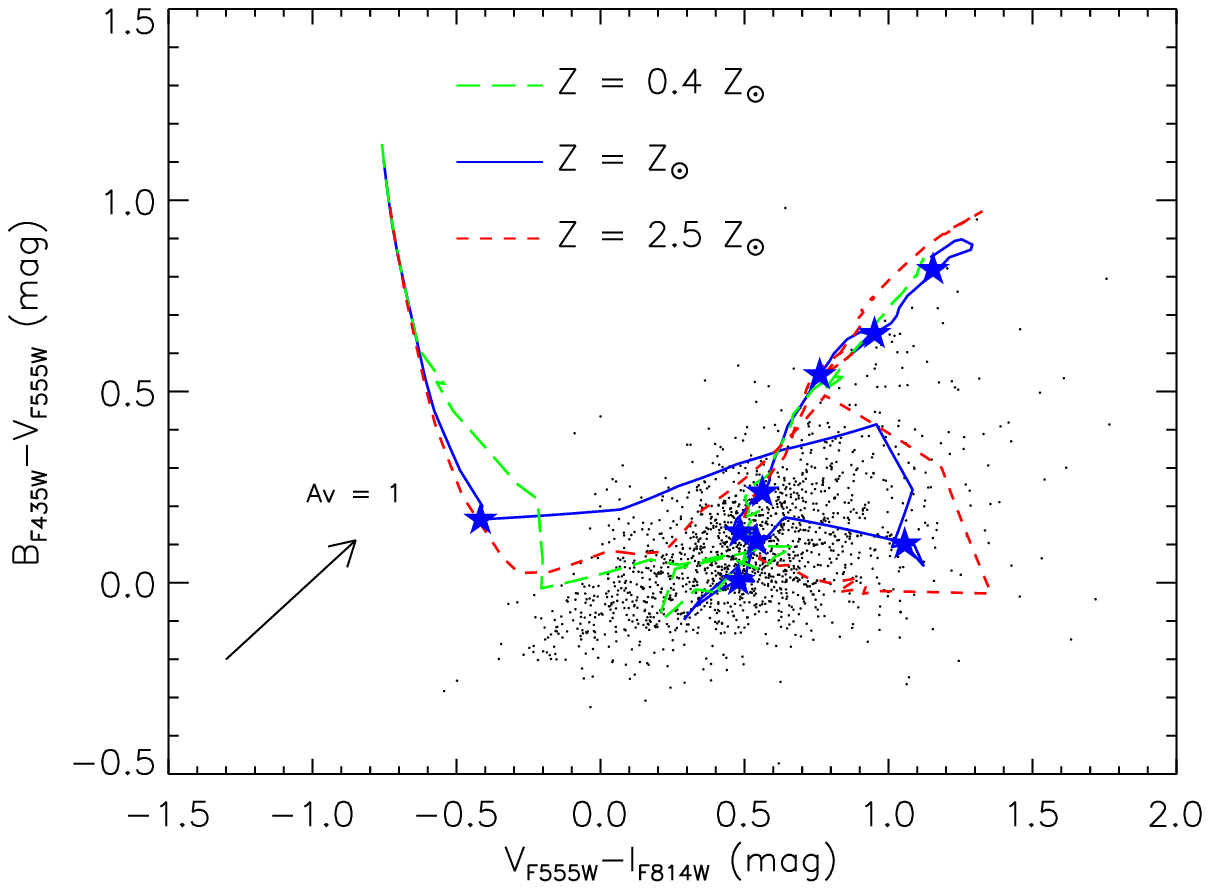} 
\caption{$\uband - \bband$ versus $\vband - \iband$ diagram
  (\emph{left}) and $\bband - \vband$ versus $\vband - \iband$ diagram
  (\emph{right}) of the 1580 clusters selected in Sect.~\ref{subsec:Selection of the different age samples} 
(not
  corrected for extinction), compared to \galev\
  SSP models using Geneva isochrones, a Kroupa IMF and three different
  metallicities, indicated in the figures. The stars indicated
  log(age/yr), starting from 6.6 (left) to 9.8 (top right) in steps of 0.4 dex.}
\label{fig:colour_vs_colour}
\end{figure*}

The \galev\ models were preferred above other SSP
models because of their freedom in stellar evolutionary models and
metallicity and because their output and treatment of extinction is
tailored to our specific \hst/\acs\ filters. Therefore, no
transformation between different photometric systems was necessary.
We used the models based on the Geneva isochrones \citep{charbonnel93, schaerer93}, which are better
suited for the age range we are interested in (i.e. age
$\lesssim100$~Myr) than e.g. Padova isochrones \citep{bertelli94, girardi00}, due to a higher time
resolution in this age range.
However, in Sect.~\ref{subsec:Systematic uncertainties} we will discuss the impact of the different isochrones on the derived cluster ages.
For the models we further adopted a \citet{kroupa01a} stellar initial
mass function (IMF) down to $0.1\msun$ and a metallicity between
$0.4 \zsun$ and $2.5 \zsun$. This range in metallicity covers the observed
metallicity of \ion{H}{II} regions in the disc of M~51 (solar to twice solar,
\citealt{diaz91, hill97}) and the slightly lower metallicity expected for older clusters. 
The extinction was varied between $0\leq \ebv \leq1$ in steps of 0.05 and the \citet{cardelli89} extinction law was applied.
Absolute magnitudes of all the clusters were calculated assuming $m-M
= 29.62$ (i.e. a distance to M~51 of $8.4\pm0.6$~Mpc,
\citealt{feldmeier97}). In Fig.~\ref{fig:colour_vs_colour} we compare
the photometry of our clusters to the SSP models in two
colour-colour diagrams.

\subsection{Systematic uncertainties}
\label{subsec:Systematic uncertainties}

\begin{figure*}
\centering
 \includegraphics[width=170mm]{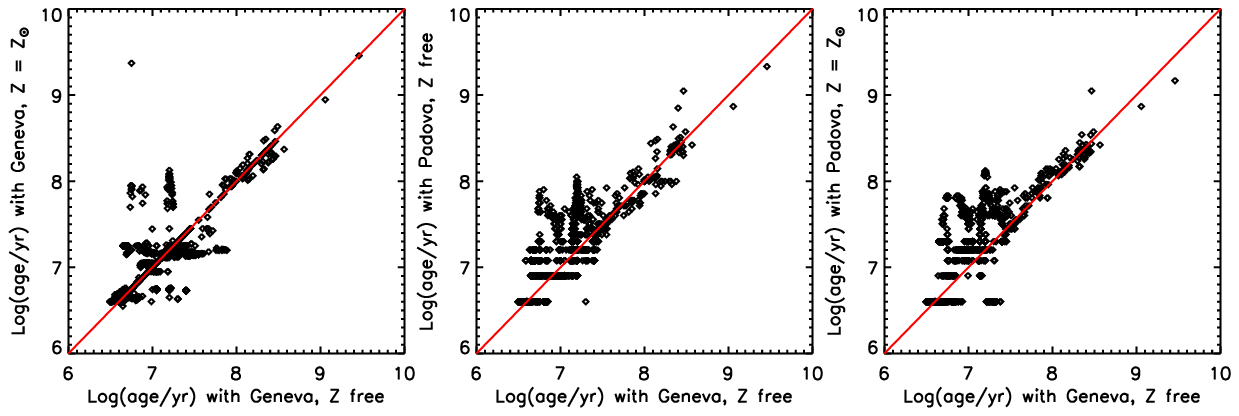} 
\caption{Comparison of cluster ages determined using different \galev\ SSP model parameters, for the 1580 clusters selected in Sect.~\ref{subsec:Selection of the different age samples}. The line shows the one-to-one correlation. {\bf Left:} Geneva isochrones for solar metallicity versus Geneva isochrones for $\zsun<Z<2.5\zsun$. {\bf Middle:} Padova isochrones versus Geneva isochrones, both for $\zsun<Z<2.5\zsun$. {\bf Right:} Padova isochrones for solar metallicity versus Geneva isochrones for $\zsun<Z<2.5\zsun$.}
\label{fig:age_vs_age}
\end{figure*}

Due to either the discreteness of the model isochrones or the rapid colour evolution of the models at certain ages in combination with the photometric uncertainties of the data, some fitting artefacts, or ``chimney's'' will generally appear in the log(mass) vs.\ log(age) plane of the clusters \citep{bastian05b, gieles05}. We find that these artefacts depend on the SSP model parameters such as the adopted isochrones and the range in metallicity. In Fig.~\ref{fig:age_vs_age} we compare the ages of clusters fitted with our adopted parameters (Geneva isochrones, Z free between $0.4\zsun$ -- $2.5\zsun$) to the ages fitted with metallicity fixed to solar or using Padova isochrones. 

First of all, Fig.~\ref{fig:age_vs_age} (left) shows that restricting the metallicity to be solar, leads to a more discrete distribution of ages and the exchange of clusters between the different chimney's: many clusters from the $\log(\mathrm{age/yr}) =7.2$ chimney (on the x-axis) are now fitted with log(age/yr) between 7.6--8 or around 6.7 (on the y-axis), i.e. they show a vertical scatter. Similarly, many clusters from the range $\log(\mathrm{age/yr})= 6.7$--8.0 (x-axis) end up in the $\log(\mathrm{age/yr}) =7.2$ chimney (y-axis), representing the horizontal scatter. 

Secondly, Fig.~\ref{fig:age_vs_age} (middle \& right) shows that with the Padova isochrones we generally fit older cluster ages for $\log(\mathrm{age}) \la 7.5$ (x-axis). These panels also show the limited time resolution of the \galev\ models based on Padova isochrones, in the first $\sim16$~Myr. This is caused by the fact that in the \galev\ models the resolution can only be as high as the lowest possible age, which is 4~Myr when using the Padova isochrones. The preference of Geneva based models for younger ages is likely a combination of a stronger contribution from red supergiants, leading to redder colours, and the higher age resolution, which lead to less ``smoothing'' of the colour fluctuations compared to the Padova based models.

We conclude from Fig.~\ref{fig:age_vs_age} that cluster age determinations based on UBVI broad-band photometry can be very uncertain in the first $\sim100$~Myr due to the uncertainties in the adopted model parameters. These uncertainties will translate into uncertainties in the different age samples, which we will select in Sect.~\ref{subsec:Selection of the different age samples}.
Although we will use the ages based on the models using Geneva isochrones and $0.4\zsun<Z<2.5\zsun$ as our ``standard ages'' in the remainder of this work, we will check, when necessary, how robust our results are against using the other models of Fig.~\ref{fig:age_vs_age}.

\subsection{Selection of the different age samples}
  \label{subsec:Selection of the different age samples}

After fitting the SSP models to our sample of 5502 clusters with $UBVI$
photometry, we applied the following selection criteria to reject
clusters with unreliable fits and to minimize incompleteness effects:
\begin{enumerate}
\item Photometry in \bband, \vband\ and \iband\ brighter than the 90\%
  magnitude completeness limits determined by S07 for a source with an
  effective (half-light) radius of 3~pc in a high background region
  (respectively 24.2, 23.8 and 22.7 mag). 
\item Photometric accuracy in \bband, \vband, \iband\ and
  \uband\ better than 0.2~mag. 
\end{enumerate} 

These two criteria reduced our cluster sample from 5502 to 2089 clusters. Since
the \uband\ imaging was not part of the source detection, we do not
have a well defined completeness limit in the \uband\ passband. We
can, however, make a rough estimate of the 90\% completeness limit in
\uband, since we performed the \uband\ aperture photometry at the
coordinates of all the resolved clusters on the \bband\ image. This
photometry was followed by the $\sigma \uband < 0.2$~mag
criterion. Using bins of $0.2$~mag wide, we measured that the
magnitude at which this $\sigma \uband$ criterion recovers $\sim$90\%
of the clusters from the total sample is $\uband = 22.0$ (see also
Fig.~\ref{fig:dU_vs_U}). This magnitude can be considered as a rough
estimate of the \uband\ 90\% completeness limit, since in a real
completeness test the recovered fraction is measured as a function of
the \emph{input} magnitude and not of the \emph{measured} (possibly
extincted) magnitude. Also, in a proper completeness test the recovered clusters are compared to an input sample which is (by definition) 100\% complete.  In our case we can only measure the effect of the $\sigma \uband$ criterion on the detected sample, which is probably already affected by some incompleteness due to extinction.
Keeping this
in mind, we used the estimated \uband\ limit
as an extra criterion to select our cluster samples:
\begin{enumerate}
\item[3.] $\uband < 22.0$~mag,
\end{enumerate}  
which further reduced our sample to 1580 clusters. 

To estimate the impact of an extinction induced bias, we can do the following: if we would assume that $\uband=22.0$~mag is the detection limit for extinction \emph{free} sources, we could  correct the measured \uband\ photometry of every source according to their best-fitting extinction, before applying the magnitude cut. This would include another 271 clusters in our sample. Still, such a correction would not account for the most extincted sources, which are not detected in the first place. This shows that extinction effects can introduce biases of the order of $\sim15$\% of the sample size. Unfortunately we have to accept that we simply \emph{can not}
fully correct for such extinction effects by the proper choice of magnitude cuts.

\begin{figure}
\centering
 \includegraphics[width=85mm]{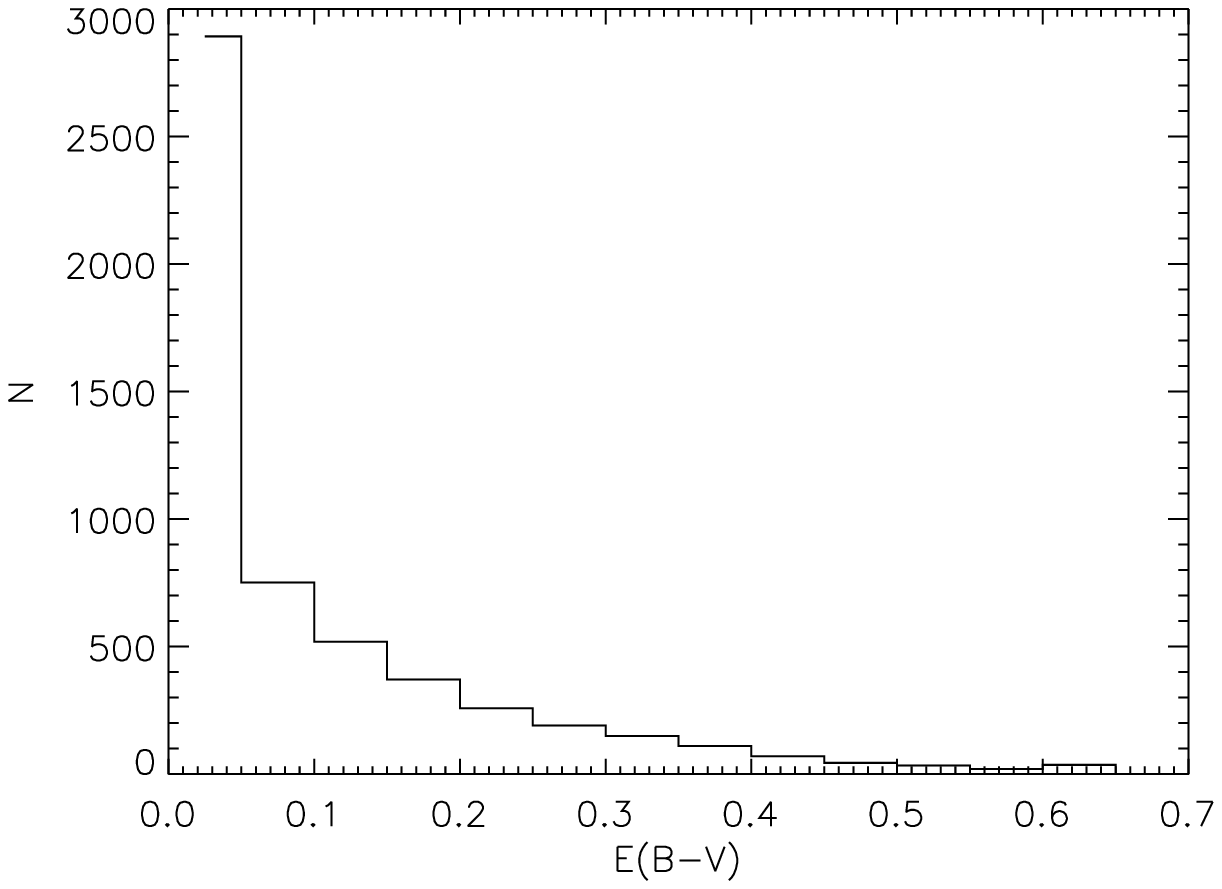} 
 \caption{Distribution of the best-fitting extinctions (\ebv) of the 5502 clusters for which we have UBVI photometry.}
\label{fig:extinction distribution}
\end{figure}

The extinction distribution is shown in Fig.~\ref{fig:extinction distribution} and is in excellent agreement with the results of \citet{bik03}. Since the distribution is peaked towards low extinctions, most of the extra clusters, if we could correct for extinction, would be located close to the \uband\ detection limit.
We therefore note that, although one can use 90\% completeness limits, since extinction (and also crowding, see e.g. S07) can not properly be taken into account, one will always end up with a sample being less than 90\% complete. This also holds for the remainder of this work, where we nevertheless will use terms like ``90\% complete'' for the sake of simplicity.

\begin{figure}
\centering
 \includegraphics[width=85mm]{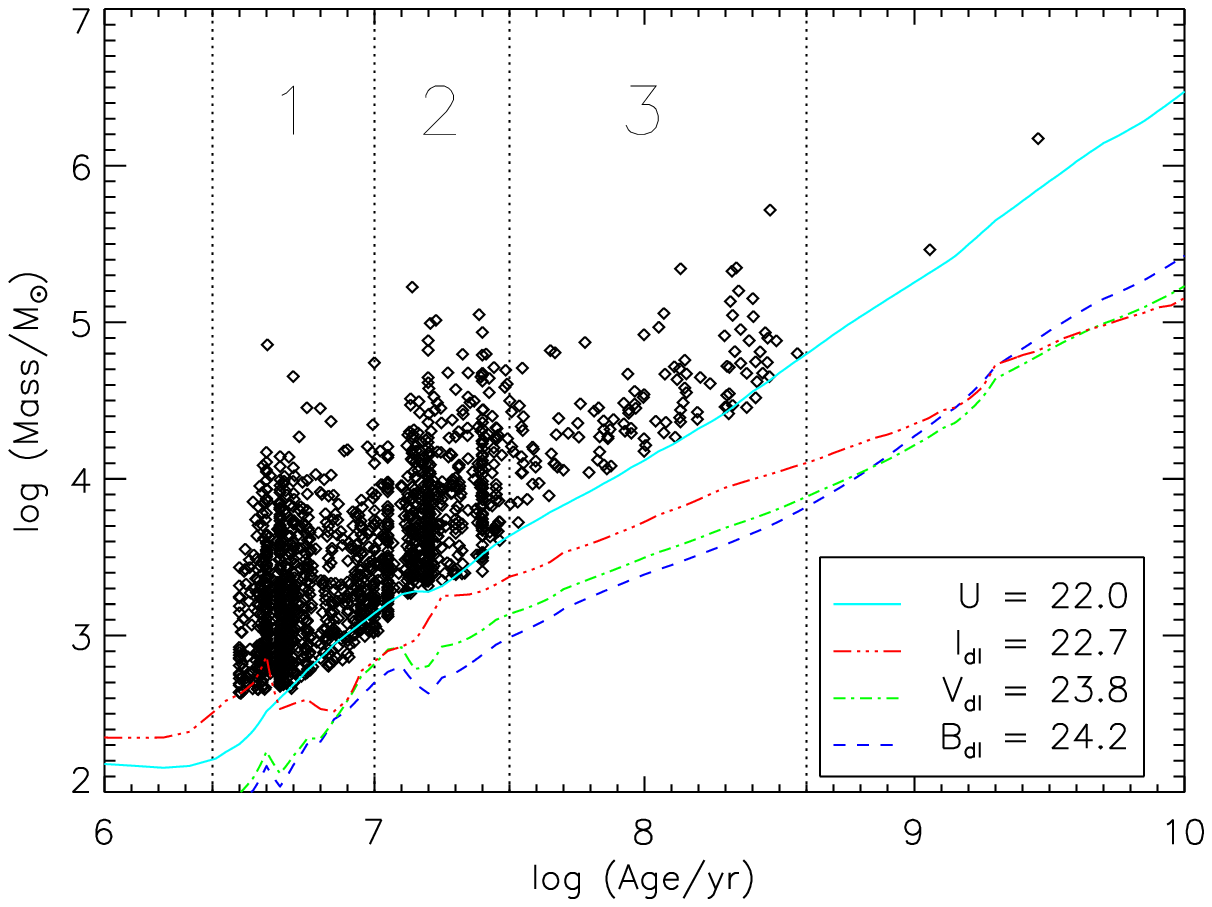} 
 \caption{Log(mass) versus log(age) diagram of the 1580 clusters with
   photometric accuracies $< 0.2$~mag and extinction corrected
   magnitudes brighter than the 90\% completeness limits in
   \bband, \vband\ and \iband\ and $\uband < 22.0$, as indicated in the lower-right
   corner. The corresponding mass limits as a function of age are
   indicated by the coloured lines. Note that the masses are
   \emph{initial} masses (i.e. corrected for stellar evolution but \emph{not} for secular evolution). The three magnitude-limited age samples are
   indicated by the numbered regions.}
\label{fig:mass_vs_age}
\end{figure}

The age-mass diagram for the clusters fullfilling criteria 1--3, is shown in
Fig.~\ref{fig:mass_vs_age}.  In this diagram two trends are
immediately visible. First, we see that the mass of the most massive
cluster generally increases with age. This is most likely a statistical
size-of-sample effect caused by the logarithmic age interval
\citep[see e.g.][]{hunter03, gieles08}. Second, we see that the mass of the
least massive clusters increases steadily with age. This is a
selection effect caused by the detection limit (i.e. the most limiting
90\% magnitude completeness limit, see criterium 1 above). For the
$BVI$ passbands we also plotted the mass evolution with age,
corresponding to the 90\% magnitude completeness limit. For the
\uband\ passband we plotted the corresponding mass evolution for the
applied $\uband = 22.0$ limit. We see that for log(age/yr)$>6.7$
\uband\ is the most limiting factor that sets the lower mass
limit. These mass evolution curves were calculated using
\galev\ models with solar metallicity. The slope of these detection
limits is caused by the evolutionary fading of the clusters: a less
massive cluster reaches a constant magnitude limit
much faster compared to a more massive cluster. These curves show that after
$\sim 20$~Myr the limiting mass in \uband\ evolves faster with age
compared to the other passbands, because the integrated flux in
\uband\ is always dominated by the hottest (and therefore fastest
evolving) stars.  We see some clusters with best-fitted masses below
the predicted limiting \uband\ mass. This is possible because of two
effects. First of all, the mass of a cluster is determined by scaling of the
\emph{complete} modelled SED (i.e. $UBVI$) and not by scaling a single
passband only. Secondly, the curve for the limiting mass is for solar
metallicity, while the best-fitting model can have a different metallicity.

We selected three \emph{magnitude-limited} cluster samples with
different ages for the remainder of this study:
\begin{itemize}
\item[$\bullet$] Sample 1:\\
$6.4 \leqslant \mathrm{log(age/yr)} < 7.0$\\
Number of clusters: 895

\item[$\bullet$] Sample 2:\\
$7.0 \leqslant \mathrm{log(age/yr)} < 7.5$\\
Number of clusters: 556

\item[$\bullet$] Sample 3:\\
$7.5 \leqslant \mathrm{log(age/yr)} < 8.6$\\
Number of clusters: 127
\end{itemize}
These three cluster samples are indicated by the numbered regions in
Fig.~\ref{fig:mass_vs_age}. The total number of clusters in these
three samples is 1578.

Due to the systematic effects described in Sect.~\ref{subsec:Systematic uncertainties}, our cluster samples will change significantly if we adopt the Padova stellar isochrones instead of the Geneva isochrones. This results in $\sim230$ clusters less in sample 1, and $\sim200$ clusters more in sample 3.  

As can be seen from Fig.~\ref{fig:mass_vs_age}, our cluster samples
have different lower-mass limits. Our youngest cluster sample contains
clusters down to estimated masses of log($M/\msun$)$\approx 2.6$, but
is 90\% complete down to $\log(M/\msun)= 3.1$, while the lowest
possible mass in our oldest cluster sample is log($M/\msun$)$\approx
3.7$. An electronic table with the photometry, ages, masses and extinctions of the clusters will be published together with a forthcoming paper \citep{scheepmaker09b}.

\section{Star formation rates, gas densities and spiral arms}
  \label{sec:Star formation rates, gas densities and spiral arms}

\subsection{Spiral arms}
  \label{subsec:Spiral arms}

In order to study the relation between star and cluster formation and
spiral structure, we first defined the two spiral arms in M~51 using
data from the \emph{Two Micron All Sky Survey} (2MASS), which we
downloaded from the \emph{NASA/IPAC Extragalactic Database} (NED). The
background subtracted $H$-band (1.65$\mu$m) image was used, which is
part of the 2MASS \emph{Large Galaxy Atlas} \citep{jarrett03}. We used
near-infrared data to trace the spiral arms, since the near-infrared
is most sensitive to older stellar populations. This means that we are
tracing the dominant mass component in the disc of M~51, \emph{independent} of
the dominant flux in the near-UV and optical of young star clusters
and massive star formation.

\begin{figure*}
\centering
 \includegraphics[width=85mm]{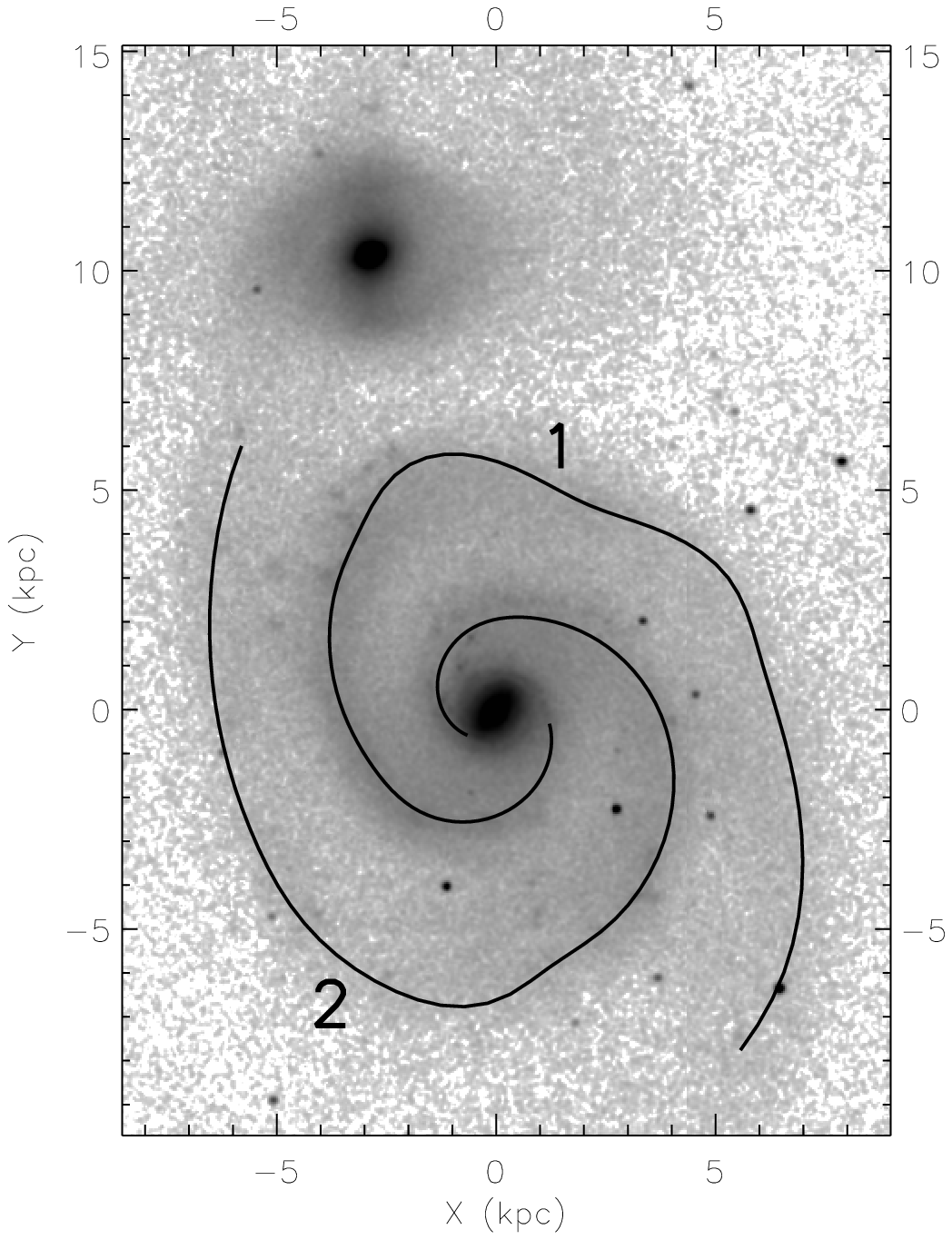}
 \includegraphics[width=85mm]{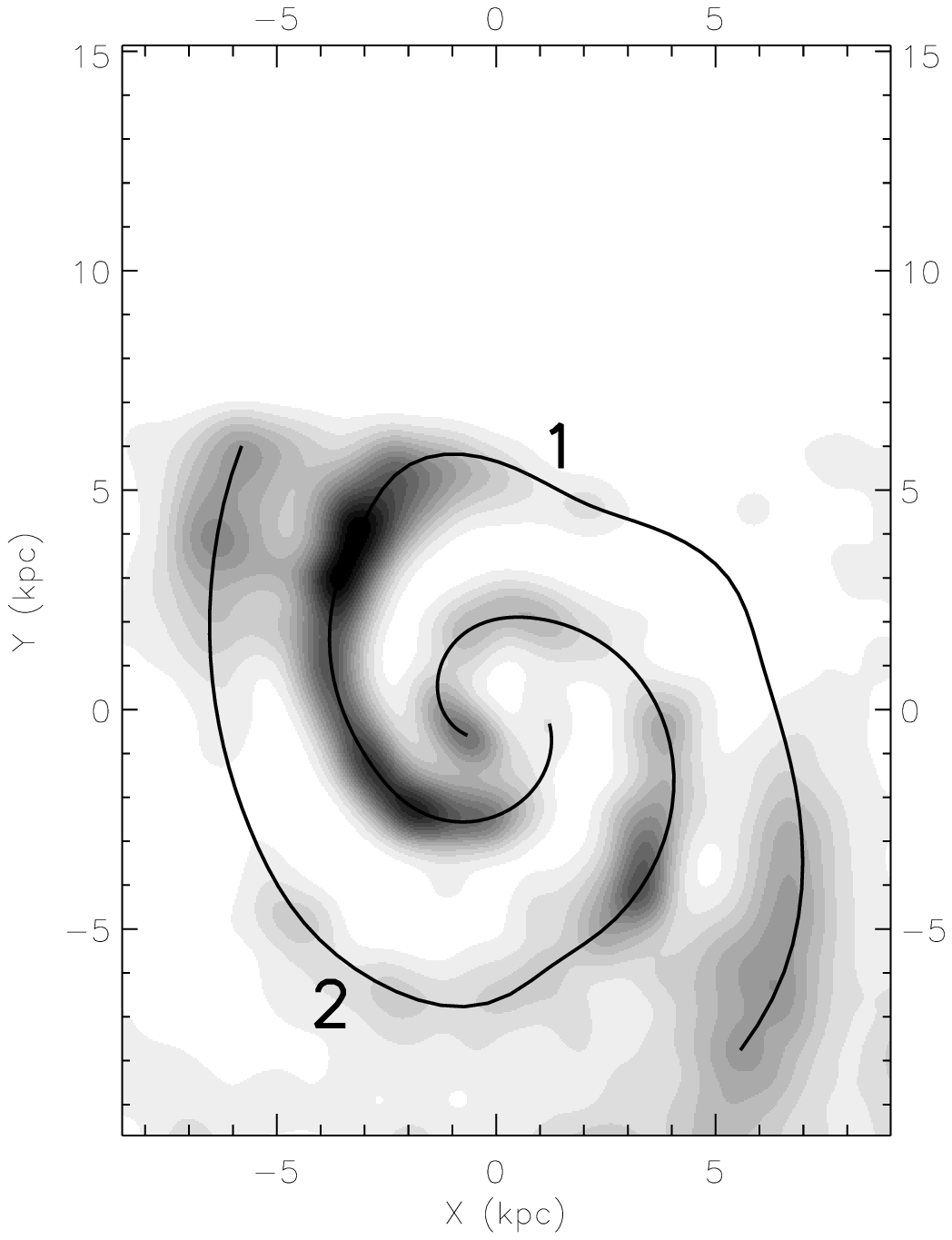}
 \caption{{\bf Left:} Location of the two spiral arms, overlaid on the
   \emph{2MASS} \emph{Large Galaxy Atlas} $H$-band image of M~51. The
   few black points are foreground stars. {\bf Right:} Same spiral
   arms, overlaid on the enhanced (masked, radial average
   subtracted and smoothed) image.}
\label{fig:spiralarms}
\end{figure*}

We enhanced the spiral structure by first masking the companion
galaxy, the inner 750~pc region, bright point sources and negative
pixels. We then subtracted radial averages and convolved the resulting
image with a Gaussian kernel with a FWHM of $25\arcsec$ ($\approx
1~kpc$). In the resulting smoothed image we identified 23 peaks
tracing the two spiral arms using \iraf/\daophot, and transformed
their positions into polar coordinates. Finally, the regions in
between the peaks were interpolated using cubic splines of the form
$\ln(R(\theta))$. The resulting interpolated spiral arms are shown in
Fig.~\ref{fig:spiralarms}, overplotted on the original background
subtracted $H$-band image (left) and the enhanced image (right). We
will refer to the arm extending towards the south-west
(i.e. bottom-right) as ``Spiral arm 1'', and to the arm extending
towards the nort-east (top-left) as ``Spiral arm 2'', as indicated in
the figure.

\subsection{Star formation rates and gas surface densities}
 \label{subsec:Star formation rates and gas surface densities}

In a recent paper \citet[][from here on referred to as
  ``K07'']{kennicutt07} studied the locally resolved SFRs and gas
densities in M~51, based on \halpha, \palpha, CO, \ion{H}{I} and \emph{Spitzer
  Space Telescope} 24$\mu$m imaging. For a series of 257 circular
apertures with a diameter of $13\arcsec$ (corresponding to
$\sim$530~pc) and distributed across the disc of M~51, these authors
present extinction corrected SFR surface densities, as well as
hydrogen gas and total gas surface densities. The SFRs are derived
from the \halpha\ line emission (after extinction correction using \palpha) using the transformation from
\citet{kennicutt98a}, and are therefore tracing the \emph{current}
star formation up to $\sim$10~Myr \citep{calzetti05}.
We used Table~1 of K07 to retrieve their list of
aperture positions and the corresponding SFR surface densities and
total gas surface densities.

\begin{figure*}
\centering
 \includegraphics[width=85mm]{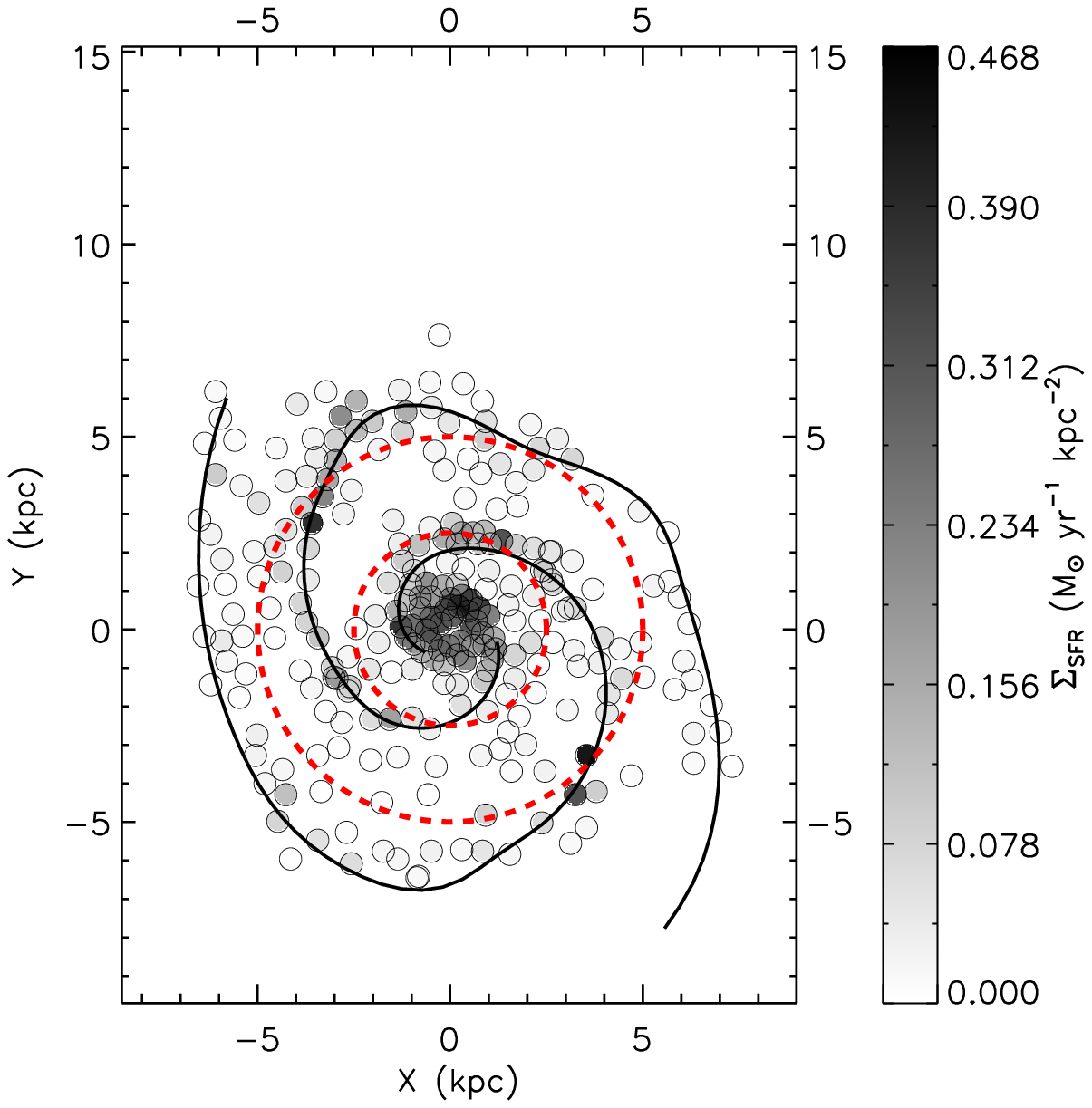} 
 \includegraphics[width=85mm]{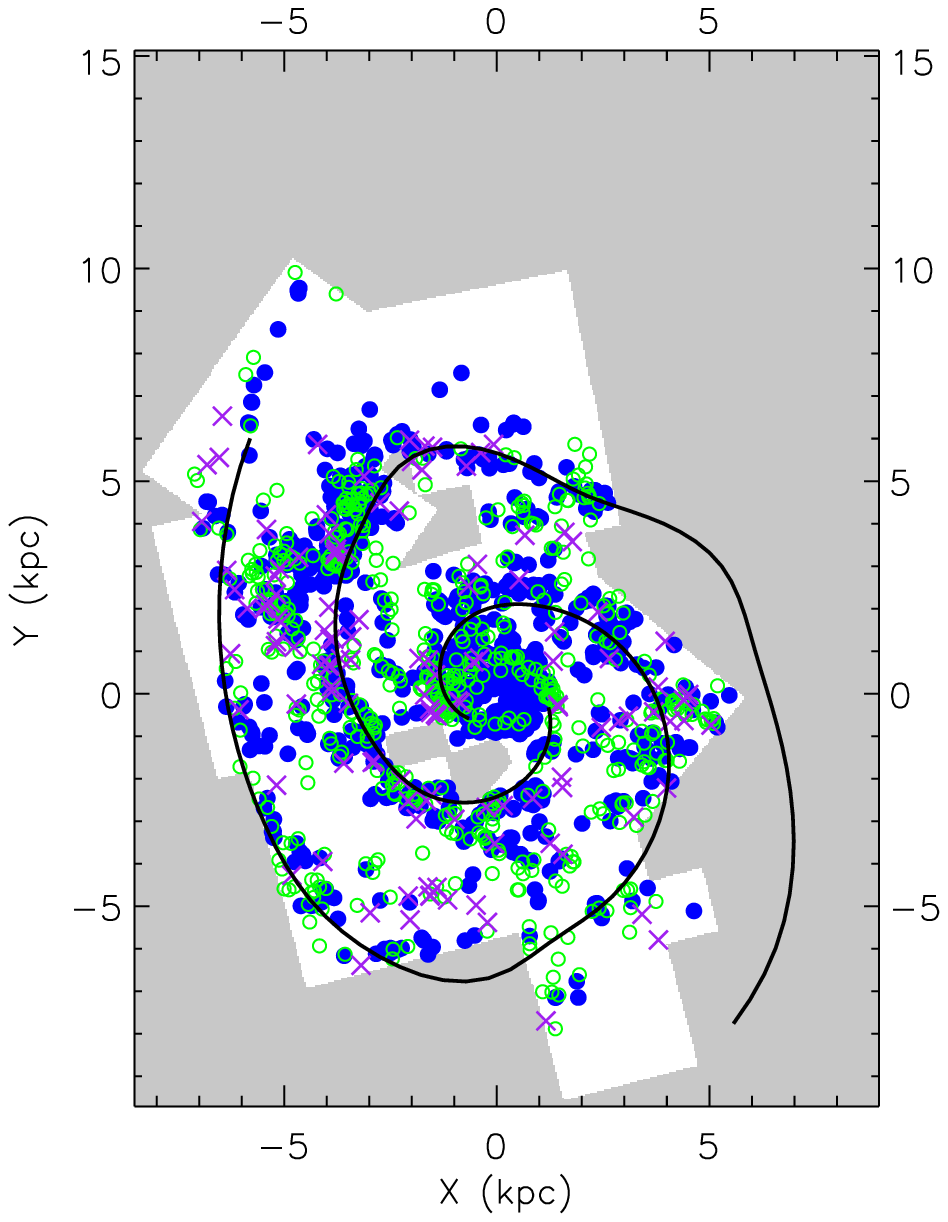} 
 \caption{{\bf Left:} Locations of the \citet{kennicutt07} $13\arcsec$
   apertures used to trace the SFR and gas surface density in
   M~51. The greyscale indicates the mean SFR surface density. The
   spiral arms from Fig~\ref{fig:spiralarms} are overplotted for
   reference. The two dashed circles are centred on the nucleus of
   M~51 and have a radius of 2.5 and 5~kpc, respectively. {\bf Right:}
   Distribution of the cluster samples with different ages across the
   disc of M~51, in the same scale as the figure on the left. Filled
   blue (greyscale: black) circles: clusters with log(age/yr)$<7.0$. Open green
   (light grey) circles: clusters with $7.0
   \leqslant$~log(age/yr)~$<7.5$. Purple (dark grey) crosses: clusters
   with $7.5 \leqslant$~log(age/yr)~$<8.6$. The regions not covered by
   the \wfpc\ fields, i.e. for which we have no age estimation of the
   clusters, are indicated in grey. The spiral arms from
   Fig~\ref{fig:spiralarms} are overplotted for reference.}
\label{fig:distributions}
\end{figure*}

The aperture positions were transformed to the frame of the
\acs\ \vband\ image using the \iraf\ task {\sc rd2xy}. In
Fig.~\ref{fig:distributions} (left) we show the resulting positions of the $13\arcsec$ apertures. The mean 
SFR surface density in every aperture is indicated in
greyscale. In the regions where two or more apertures are overlapping
the average value was used. 

\section{Spatial distribution of star clusters and star formation}                
  \label{sec:Spatial distribution of star clusters and star formation}

In Fig.~\ref{fig:distributions} (right) we show the 2-dimensional
spatial distribution of the three cluster samples in M~51. Also shown
are the interpolated spiral arms from Fig.~\ref{fig:spiralarms} and
the regions (white) which are covered by the \wfpc\ pointings,
i.e. the regions for which we have derived ages of the clusters.

We can see from Fig.~\ref{fig:distributions} that most current star formation is taking place in the centre of M~51 and at a galactocentric radius of $\sim5$~kpc, and that these are also the locations where most of the young star clusters reside. We study these correlations more closely in Sect.~\ref{subsec:Radial distribution} with the help of radial distributions.

Fig.~\ref{fig:distributions} shows clearly that the majority of the
clusters is closely tracing the spiral arms. This is expected, since
the majority of the clusters are believed to be formed by the passing
spiral density wave. Older clusters are expected to trace the spiral
arms less closely, and we indeed see that the clusters from sample 3
($\mathrm{log(age/yr)} \geqslant 7.5$) are more uniformly distributed.
This is consistent with the result of \citet{hwang08}, who found that,
using only the \acs\ $BVI$ data, the bluest clusters in M~51 are
closely associated with the spiral structure and redder clusters are
more uniformly distributed. We look at this in more detail in Sect.~\ref{subsec:Azimuthal distributions} using azimuthal distance distributions.

East (left) of the centre of M~51 the clusters are more concentrated
on the convex (outward) side of spiral arm 1 compared to the concave
(inward) side, up to $\sim$5~kpc. For spiral arm 2 in the south-west
this effect is less obvious, although we note that for spiral arm 2 we
are much more hampered by selection effects due to the incomplete
\wfpc\ coverage.

Fig.~\ref{fig:distributions} also shows that the distribution of
clusters itself is ``clustered'' in so-called cluster complexes (see
e.g.~\citealt{efremov95, bastian05a}). These complexes are the largest groups in
the hierarchy of star formation, which originates from the fractal
distribution of the interstellar gas \citep{elmegreen96b, efremov98,
  elmegreen01b}. In such a hierarchy, the smallest observed structures are
the youngest ones. This trend is visible in
Fig.~\ref{fig:distributions}, were the spatial distribution of the
youngest cluster sample shows most clustering and the amount of
clustering decreases with the age of the cluster sample. We will
quantify this more precisely in Sect.~\ref{subsec:Auto-correlation
  functions} with the help of two-point autocorrelation functions.

\subsection{Radial distributions}
  \label{subsec:Radial distribution}

\begin{figure}
\centering
 \includegraphics[width=85mm]{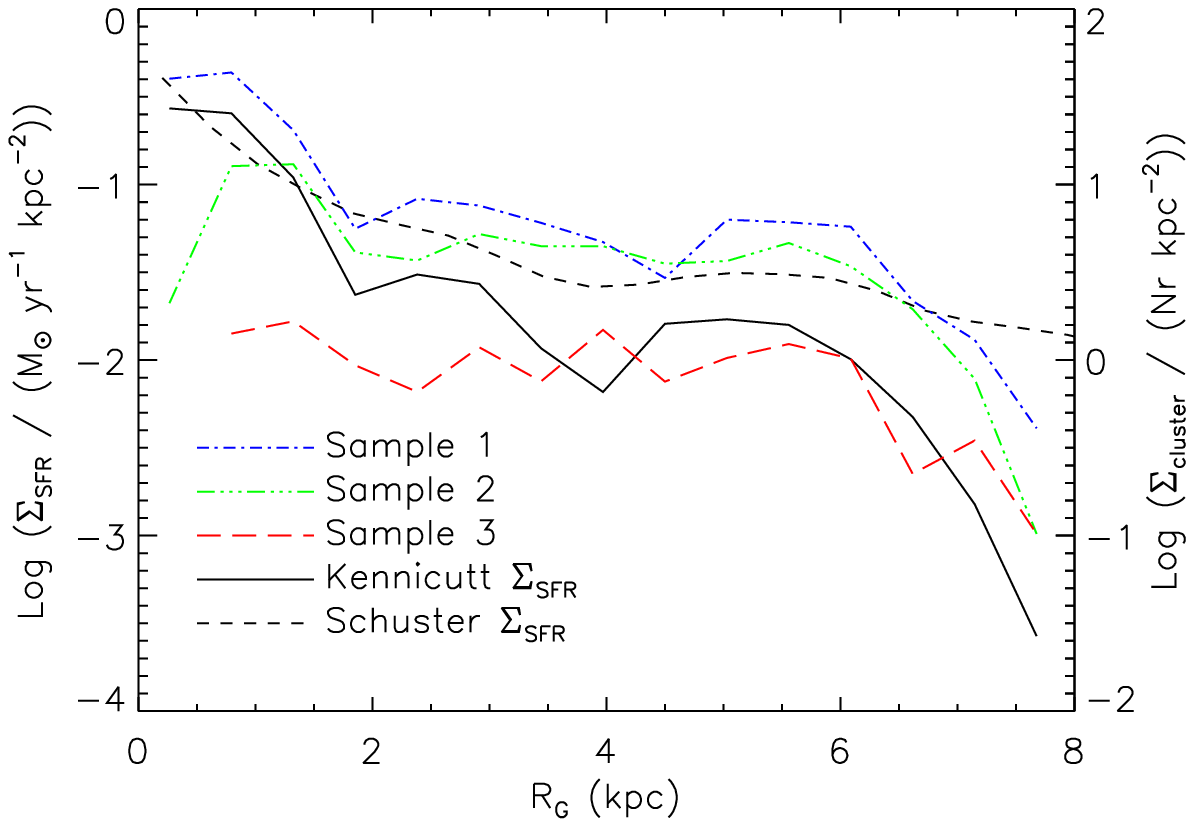} 
\caption{Radial number density distributions of the different cluster
  samples (right axis), compared to the radial distribution of the
  SFR surface density from \citet{kennicutt07} and \citet{schuster07}
  (left axis).}
\label{fig:radial distribution}
\end{figure}

\begin{figure}
\centering
 \includegraphics[width=85mm]{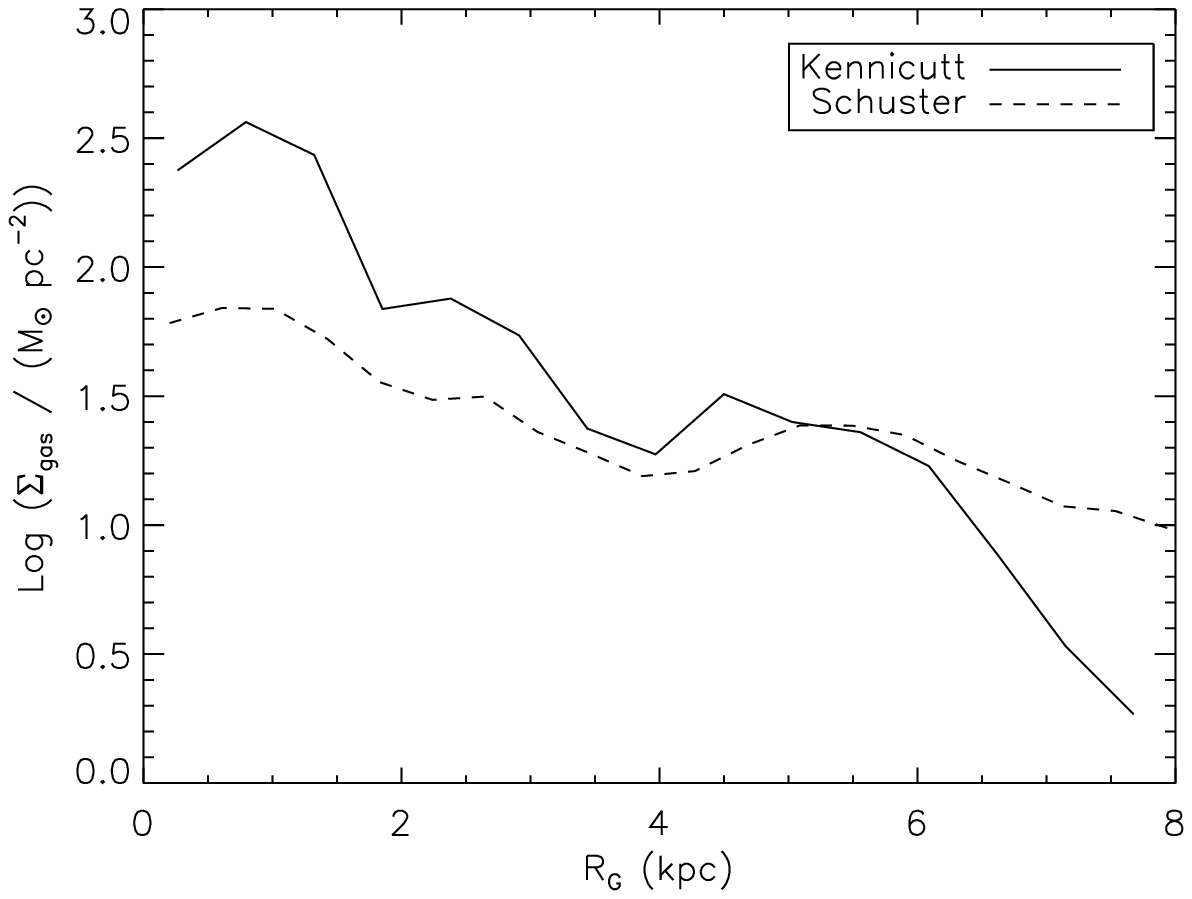} 
\caption{Radial distributions of the total gas (i.e. including helium
  and metals) surface density. The solid curve is derived from the
  data of \citet{kennicutt07}, while the
dashed curve is from Fig.~4 of \citet{schuster07}.}
\label{fig:radial gas}
\end{figure}

\begin{figure}
\centering
 \includegraphics[width=85mm]{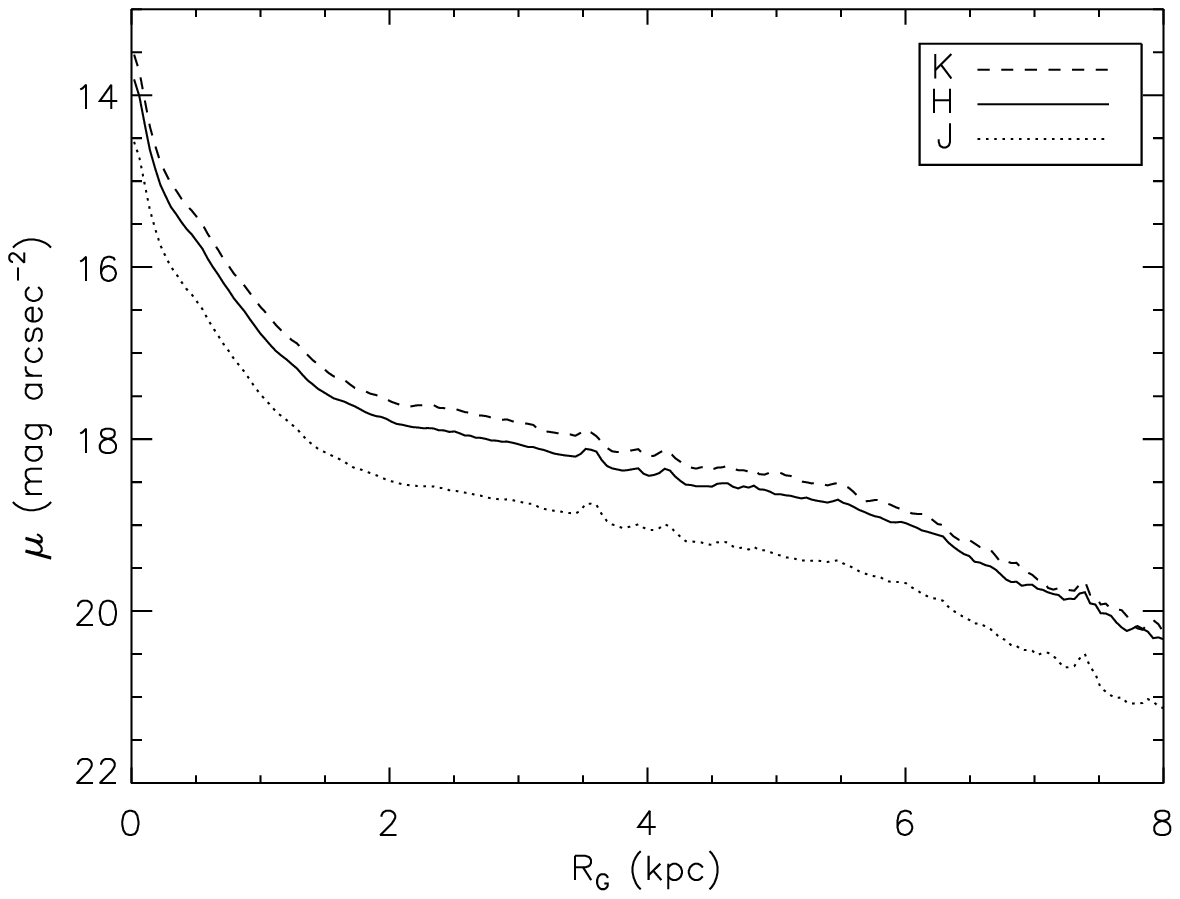} 
\caption{Surface brightness distribution of M~51 in the 2MASS $J$, $H$
  and $K$-band.}
\label{fig:radial 2mass}
\end{figure}

From the 2-dimensional spatial distributions of SFR, gas surface
densities and star clusters (Fig.~\ref{fig:distributions}) we derived
radial density distributions using $13\arcsec$ wide circular annuli
and, in case of the clusters, masking out the regions not covered by
the \wfpc\ fields (i.e. the grey area in Fig.~\ref{fig:distributions}
(right)). The SFR and gas apertures cover the entire field of view, so
no masking was necessary for those. In Fig.~\ref{fig:radial
  distribution} we compare the radial number density distributions of
the different cluster samples directly to the SFR surface density
distribution. Fig.~\ref{fig:radial gas} shows the radial distribution
of the total gas surface density (including helium and metals), and in
Fig.~\ref{fig:radial 2mass} we show the radial surface brightness
distributions of the older stellar populations (no masking applied),
derived from 2MASS $J$, $H$ and $K$-band imaging \citep{jarrett03}. In
a recent paper, \citet{schuster07} also presented radial distributions
of SFR and gas density in M~51 using 20~cm radio continuum data for
the SFR density and a combination of CO and \ion{H}{I} data for the gas
density.  These distributions can be used as an independent check to
the data taken from K07 and are also shown in Figs.~\ref{fig:radial
  distribution}~\&~\ref{fig:radial gas}.  We note that
\citet{schuster07} have corrected their distributions (i.e. the
galactocentric radii) for the inclination angle of the disc of M~51,
while the cluster densities and the densities from K07 have only been
divided by an additional factor of 1.07 to correct for the effect of
the inclination on the projected area of the apertures/annuli. With an
inclination angle of only $20\,^{\circ}$ \citep{tully74}, however,
this difference is negligible for our purposes. Figures~\ref{fig:radial distribution}~\&~\ref{fig:radial gas} show that the \citet{schuster07} distributions are smoother than the K07 distributions. This is consistent with the radio data having a lower resolution (namely a half power beamwidth of $15\arcsec$ for the SFR data and $13\arcsec$ for the gas data, compared to the $13\arcsec$ apertures of K07). The offset in the gas surface densities (Fig.~\ref{fig:radial gas}) between K07 and \citet{schuster07} is probably due to differences in the conversion factors used to calculate the \htwo\ column density based on the CO data.   

From Fig.~\ref{fig:radial distribution}~\&~\ref{fig:radial gas} we see
that both the SFR and gas density are peaking in the centre, and show two
other distinct peaks at $\sim$2.5 and $\sim$5~kpc. These peaks are
\emph{not} simply a side effect of tracing a larger part of the spiral
arms at these distances, as they are not clearly visible in the
$K$-band surface brightness distribution (Fig.~\ref{fig:radial
  2mass}), while this band was used to define the spiral arms
(Sect.~\ref{subsec:Spiral arms}). The locations of the peaks however,
coincide with the locations of the four most active star forming
regions in the spiral arms of M~51, as can be seen in
Fig.~\ref{fig:distributions} (left).

We see from Fig.~\ref{fig:radial distribution} that the radial
distribution of clusters younger than 10~Myr (sample~1) follows the
SFR density distribution to a high degree (including the 2
peaks). Sample 2 shows a much flatter distribution but is still
peaking towards the centre, followed by a dip within the inner
$\sim$1~kpc, while the oldest cluster sample (sample~3) shows a flat
distribution. These conclusions hold if we would use the Padova isochrones, and it agrees very well with the results of
\citet{hwang08}, who find peaks at $\sim$3 and $\sim$6~kpc in the
radial distribution of their ``class 2'' clusters (i.e. clusters which
are more distorted and/or have multiple neighbours) and an
increasing flattening of the radial distribution with increasing
cluster colour (i.e. with increasing age, although some of their
reddest clusters might be very young and extincted).

The peaks in $\Sigma_{\mathrm{SFR}}$, $\Sigma_{\mathrm{gas}}$ and
$\Sigma_{\mathrm{cluster}}$ at $R_{\mathrm{G}} \approx 5$~kpc are
consistent with an enhanced star formation activity caused by
corotation \citep{elmegreen89b}, since the corotation radius of M~51
is at $R_{\mathrm{G}} = 5.4$~kpc, assuming a circular
velocity of $V = 200\kms$ \citep{garcia-burillo93} and a pattern speed
of $\Omega_{\mathrm{p}} = 37\kms \mathrm{kpc}^{-1}$ \citep{zimmer04}. The peaks at $R_{\mathrm{G}} \approx
2.5$~kpc could be explained by the presence of the 4:1 resonance,
which is were $\Omega - \kappa/4$ equals the pattern speed, $\Omega$
being the angular velocity of the stars and interstellar material and
$\kappa$ the epicyclic frequency \citep{contopoulos86,
  contopoulos88}. \citet{elmegreen89b} have found evidence for a
spiral arm amplitude minimum at this resonance, consistent with the
gaps, orbital loops and stellar orbit crowding, that are expected for
this resonance \citep{contopoulos86}. It is also suggested that orbit
crowding and loops result in more gas shocking. This would lead to
enhanced star formation
at the 4:1 resonance radius, resulting in an enhancement of
$\Sigma_{\mathrm{SFR}}$ at $\sim$2.5~kpc, which is then followed by an
enhanced $\Sigma_{\mathrm{cluster}}$ for the youngest cluster sample.

\subsection{Azimuthal distributions}
 \label{subsec:Azimuthal distributions}

\begin{figure}
\centering
 \includegraphics[width=85mm]{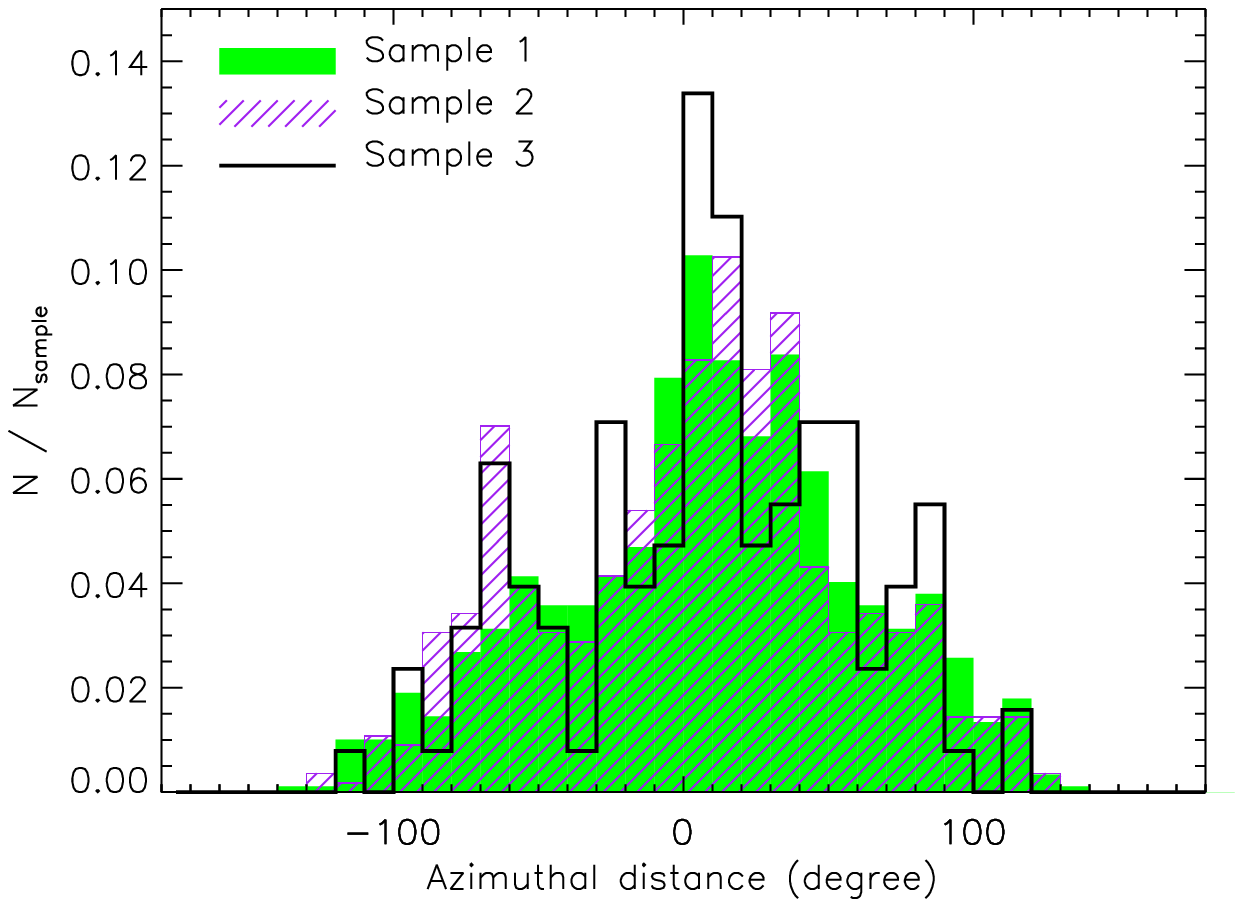} 
\caption{Normalized distribution of the azimuthal distance of the clusters to their closest spiral arm.}
\label{fig:azimuthal}
\end{figure}

If the formation of star clusters is triggered by gas shocking due to the passing spiral density wave \citep{roberts69}, we expect the youngest star clusters to be located closer to the spiral arms compared to older clusters. Using the spiral arms from Fig.~\ref{fig:spiralarms}, we derived for every cluster their azimuthal distance from their closest spiral arm, assuming circular orbits. No corrections for the position
angle (PA) and inclination angle ($i$) were applied, since these will
practically be negligible for the almost face-on orientation of M~51
($\mathrm{PA} = 170\degr$, $i = 20\degr$, \citealt{tully74}). In Fig.~\ref{fig:azimuthal} we show the distribution of the azimuthal distances for the three cluster samples (sometimes referred to as the ``circular
distribution'', e.g.~\citet{boeshaar77}), normalized by the total number of clusters in every sample. The figure shows that \emph{all} cluster samples are peaking at the location of the spiral arms (i.e. an azimuthal distance of $0\degr$). Although the youngest clusters (sample 1) show the most smooth distribution (i.e. with less scatter) compared to the other two samples, K-S tests do \emph{not} indicate statistically significant differences between the samples.
This result is robust against the variations of the SSP model parameters, used to derive the cluster ages, as described in Sect.~\ref{subsec:Systematic  uncertainties}.

\begin{figure}
\centering
 \includegraphics[width=85mm]{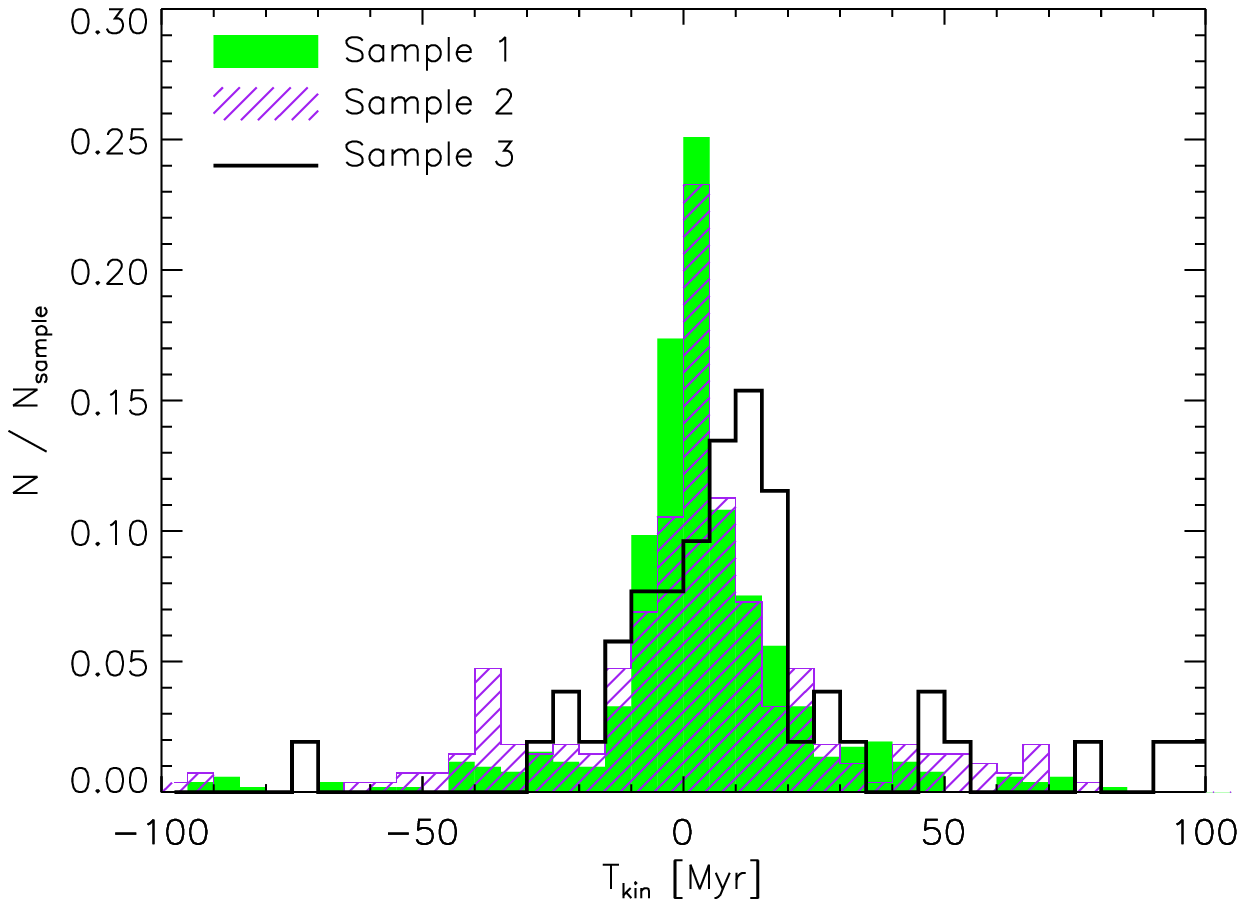} 
\caption{Normalized distribution of the kinematic ages of the clusters from their closest spiral arm, for the clusters with a galactocentric distance $< 4$~kpc.}
\label{fig:azimuthal_tkin}
\end{figure}

The fact that the oldest cluster sample is peaking at the centre of the spiral arms is likely a side effect of the enhanced cluster formation around the co-rotation radius (Sect.~\ref{subsec:Radial distribution}). Around the co-rotation radius the drift velocity away from the spiral arms is very low, meaning that older clusters once formed in these regions are still located close to the arms.
The azimuthal distance of the clusters from the spiral arms expressed in terms of ``kinematic age'' will therefore be a better description of the differences in the circular distributions between the three cluster samples. In Fig.~\ref{fig:azimuthal_tkin} we show the circular distributions in units of kinematic age, \tkin, which is the azimuthal distance to the closest spiral arm, divided by the local drift velocity at the location ($R_{\mathrm{G}}$) of the cluster. A negative \tkin\ implies that the cluster is still approaching its closest spiral arm. In Fig.~\ref{fig:azimuthal_tkin} we only plotted clusters with $R_{\mathrm{G}} < 4$~kpc, i.e. well within the co-rotation radius, since close to co-rotation \tkin\ will diverge and outside the co-rotation radius the ``kinks'' or near-circularity of the spiral arms could make the intersection of the cluster orbit and spiral arm ambiguous. Note that this prescription for \tkin\ does not account for the passing of multiple spiral arm, e.g. an old cluster which has passed numerous spiral arms in its orbit can still be appointed a very low kinematic age, if it currently happens to be located close to an arm. 

This notwithstanding, we see in Fig.~\ref{fig:azimuthal_tkin} that the oldest cluster sample is peaking at older kinematic ages (10--15~Myr) compared to the younger two cluster samples, which peak at 0--5~Myr. We also see that sample 1 has a larger fraction of clusters at the youngest kinematic ages left of the peak compared to sample 2. The kinematic ages of the peaks in the distributions are younger than the typical ages of the clusters in the samples, indicating that the majority of the clusters formed $\sim5$--20~Myr before their parental gas cloud would have reached the centre of the spiral arm. 
This is in agreement with the sharp edge of the dust clouds, seen on the \hst/\acs\ image (see e.g. the dust lane covered by field 1 and 2 in Fig.~\ref{fig:footprints}). For a collapse time of the order of $10^{6}$ years we expect most contracting clouds to have formed a cluster well before the centre of the spiral arm is reached. \citet{tamburro08} find a timescale of 3.3 Myr in M~51 for the transition from \ion{H}I emission, via CO emission to 24$\mu$m emission from current star formation. Since \ion{H}{I} first has to condense to $\mathrm{H}_{2}$, the collapse time has to be shorter than 3.3~Myr.

The circular distributions are reasonably
symmetrical around their peaks, which is consistent with the circular
distributions found for \ion{H}{II} regions in other spiral galaxies
(e.g. in NGC~4321, \citealt{anderson83}) and it is in qualitative
agreement with the density wave models which predict a more smoothed
surface brightness distribution across spiral arms
\citep[e.g.][]{bash79}, as opposed to the models which predict an asymmetric distribution \citep{roberts69}.

\subsection{Clustering of star clusters}
  \label{subsec:Auto-correlation functions}

The degree of clustering for different cluster samples can be
quantified and compared by using the two-point correlation function
\citep[][\S~45]{peebles80}. Applied to clusters from the same cluster
sample, the two-point correlation function becomes an autocorrelation
function. In the following, we broadly follow the method described in
\citet{peebles80} and \citet{zhang01}. Zhang et al. applied two-point
correlation functions to star clusters in NGC~4038/39 (the
``Antennae'').  The autocorrelation function is defined as:
%
\begin{equation} \label{eq:autocorrelation}
1+\xi (r) = \frac{1}{\bar{n}N}\sum_{i=1}^{N} n_{i}(r),
\end{equation} 
where $n_{i} (r)$ is the number density of clusters found in an
aperture of radius $r$ centred on, but excluding cluster $i$. $N$ is the total number of clusters and $\bar{n}$ is the
average number density of clusters (excluding the grey regions in
Fig.~\ref{fig:distributions} (right)). In general, $\xi (r)$ is
defined such that $\bar{n}[1+\xi(r)]\ud V$ is the probability of
finding a neighbouring cluster in a volume $\ud V$ within a distance
of $r$ from a random cluster in the sample \citep{peebles80}. For our
spatial distribution of clusters this means that $1+\xi (r)$ is a
measure for the mean surface density within radius $r$ from a cluster,
divided by the mean surface density of the total sample (i.e. the
surface density enhancement within radius $r$ w.r.t.\ the global
mean). Therefore, a random distribution of clusters will have a
constant $1+\xi(r) = 1$, while for a clustered distribution $1+\xi(r)
> 1$.

Since the star clusters in M~51 reside primarily in the disc, while
the observations cover a larger projected area, all cluster samples
will show some degree of autocorrelation and the absolute scaling of
$1+\xi(r)$ will be arbitrary. We can, however, compare the relative
scaling of $1+\xi(r)$ between the different cluster samples as well as
the different functional forms.

\begin{figure}
\centering
 \includegraphics[width=85mm]{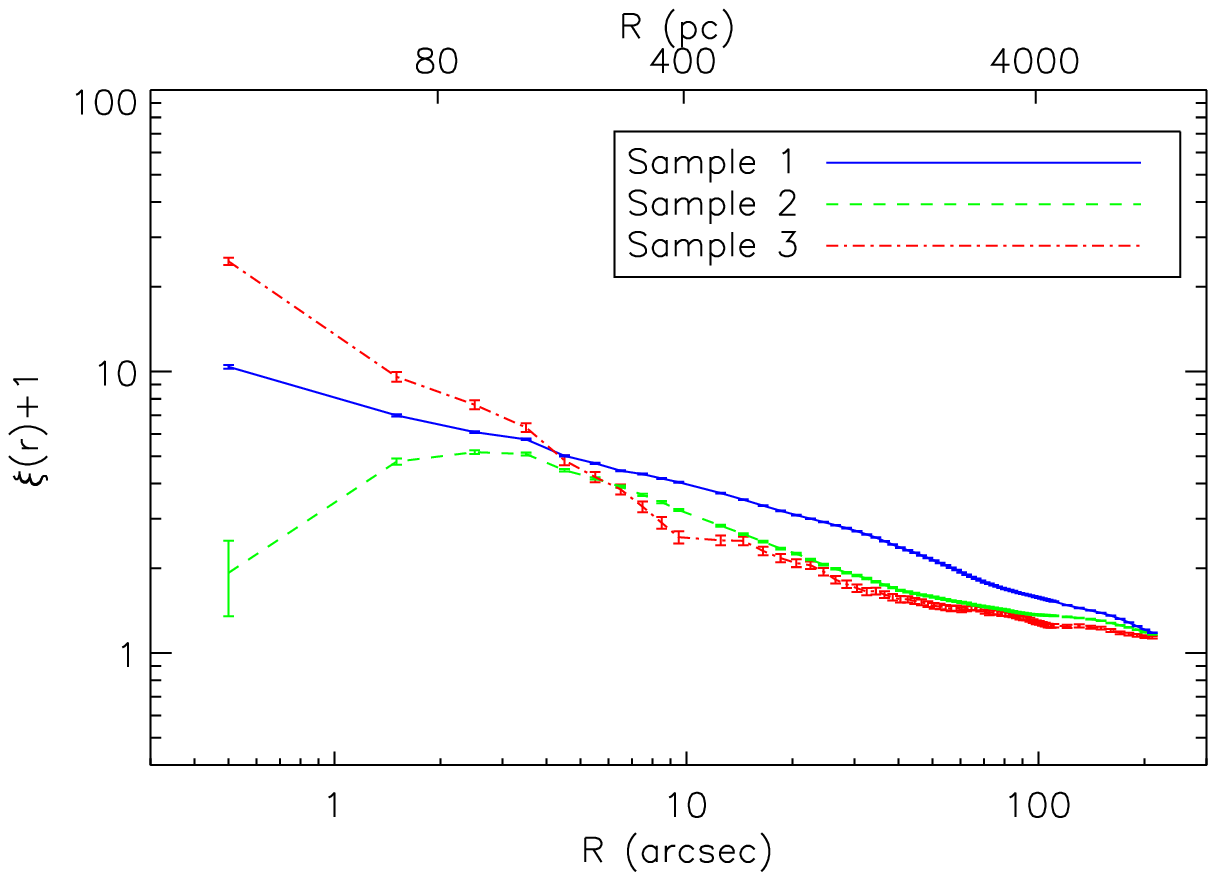} 
\caption{The autocorrelation function as function of radius for the
  three cluster samples.}
\label{fig:autocorrelation}
\end{figure}

In Fig.~\ref{fig:autocorrelation} we show the autocorrelation
functions as a function of radius for the three cluster
samples. Before the autocorrelation was calculated, the original
\acs\ coordinates were mapped onto a new image with $1\arcsec$
resolution to reduce computing time. We took into account that as a
result of this binning, one pixel could contain multiple clusters.
For the errorbars we used \citep[][\S~49]{peebles80}:
\begin{equation}
\delta(r) = \left(\frac{1}{2}\sum_{i=1}^{N} n_{\mathrm{p}}(r)\right)^{-1/2}, 
\end{equation}
where $n_{\mathrm{p}}(r)$ is the number of pairs formed with the
central cluster $i$ of the current aperture, and the factor $1/2$
accounts for not counting every pair twice.

Fig.~\ref{fig:autocorrelation} shows that for $r>4\arcsec$ ($\sim160$~pc) the
clusters with ages $<10$~Myr (sample 1) are most clustered (largest $\xi (r)$) and that
the amount of clustering decreases with age. This is consistent with
the visual interpretation of Fig.~\ref{fig:distributions}
(right). This result is robust against variations in the isochrones (Padova or Geneva) or metallicity range used in the SSP model fitting of our clusters. Below $r \sim10\arcsec$ (i.e. $\sim400$~pc) we see a steady
increase in the clustering of sample 3, surpassing sample 1 below
$r \sim4\arcsec$. We also see that below this radius the amount of
clustering for sample 2 decreases. However, if we would use the Padova isochrones we find that the autocorrelation function of sample 3 is much flatter, while the turnover of sample 2 decreases. Due to this degeneracy, we can not draw any firm conclusions on the age dependency of the autocorrelation function on scales  below
$\sim4\arcsec$.

Moreover, on these smallest scales we can be biased by selection effects in the original sample selection of
resolved sources of \citet{scheepmaker07}. In this sample selection,
sources with a neighbouring source within 5 \acs\ pixels
(i.e. 0.25\arcsec) were rejected, possibly lowering the amount of
autocorrelation in the first radius bin in
Fig.~\ref{fig:autocorrelation}. Crowding effects
and the high background in the most compact regions in which the star
clusters form, will also introduce a bias against the detection of clusters
on the smallest scales. This bias will be stronger for younger
clusters, since in a hierarchical (i.e. ``scale free'') model of star formation the
youngest structures are also the most compact ones. Correcting for such biases will increase the autocorrelation function of our youngest sample (sample 1), as we will show below.

The general
trend of $1+\xi$ decreasing with radius is expected for hierarchical (i.e. ``scale free'' or ``fractal'')
star formation \citep[e.g.][]{gomez93, larson95, bate98}. 
If we consider a compact, isolated grouping of newly formed star clusters, we expect such grouping to disperse with a typical
velocity dispersion $<10~\kms$ \citep{whitmore05}. For a 10 and 30~Myr
old population of clusters, autocorrelation is therefore expected to
increase on scales smaller than roughly 100 and 300~pc, respectively
(which are lower limits, considering the initial size $>0$~pc of the
GMC that formed the grouping). On larger scales, the autocorrelation for an isolated grouping of clusters is expected to level off. However, this will only hold if the grouping was \emph{not} part of a larger hierarchy of star formation. In a hierarchical model, on the other hand, the clusters will be correlated with other clusters from other groupings on larger scales,
together forming a group higher up in the hierarchy of star formation.
The smooth decline of $1+\xi$ with radius, observed for our cluster samples on scales $r>3\arcsec$, is therefore consistent with these populations being part of this hierarchy. 
The velocity dispersion will cause an overall decrease of the amount of clustering of the star clusters with time \citep{bate98}, consistent with Fig.~\ref{fig:autocorrelation}.

For a hierarchical, or fractal, distribution of stars or clusters, the autocorrelation function will show a powerlaw dependency with radius of the form $1+\xi (r) \propto r^{\eta}$ \citep{gomez93}. For such a distribution, the total number of clusters ($N$) within an aperture increase with radius as $N\propto r^{\eta}\cdot r^{2}  = r^{\eta+2}$, which shows that $\eta$ is related to the (two-dimensional or projected) fractal dimension $D_{2}$ as $D_{2} = \eta +2$ \citep{mandelbrot83, larson95}.
The distribution of star formation and interstellar gas over a large range of environments is observed to show a (three-dimensional) fractal dimension of $D_{3} \sim2.3$ \citep[e.g.][]{elmegreen96a, elmegreen01b}
Projecting this three-dimensional fractal on a plane will result in a two-dimensional fractal dimension  $D_{2} \approx D_{3}-1 = 1.3$ \citep{mandelbrot83, beech92, elmegreen01b}.
In the spiral galaxy NGC~628, \citet{elmegreen06} observe star-forming regions with $D_{2} = 1.5$.

If we fit powerlaws to the straight parts of the autocorrelation
functions in Fig.~\ref{fig:autocorrelation} ($4\arcsec < r <50\arcsec$), we find that the decreasing trend of sample 1 and 2 can be 
approximated with a power law index of around $-0.4$ (fractal $D_{2} = 1.6$). Sample 3 has a lower index of around $-0.8$ ($D_{2} = 1.2$), but as mentioned before, this index depends strongly on the stellar isochrones used: using the Padova isochrones, some clusters from sample 2 are fitted with older ages (see Fig.~\ref{fig:age_vs_age}) and move to sample 3, flattening the autocorrelation function to a power law with an index of around $-0.4$. 
Interestingly, an index of $-0.4$ is similar to the index \citet{zhang01} find for candidate young stars in the Antennae galaxies (namely an index of $-0.41$), while the same authors find steeper slopes for the autocorrelation functions of young (5--160~Myr) star clusters (namely indices between $-0.74$ and $-1.06$). 
In an attempt to minimize the possible bias due to crowding of the youngest clusters, we also determined the autocorrelation function of all clusters in sample 1 with $\log(M/\msun)>3.7$. This sample will have a higher completeness in the densest star forming regions. For this mass-limited sample we find a power law index of $-0.8$ ($D_{2} = 1.2$, Geneva) or $-0.7$ ($D_{2} = 1.3$, Padova), similar to the indices of \citet{zhang01} and consistent with the fractal dimension of $\sim1.3$, mentioned above. This suggest that the flatter slope of the full sample 1 population of clusters is indeed biased, and that the unbiased index of the autocorrelation function  of young clusters is similar for different galaxies. In other words, this hints towards a universal fractal dimension of hierarchical star formation. More studies in different galaxies and of different nature are necessary to strengthen this point.

\subsection{Spatial correlation between star clusters and star formation}
  \label{subsec:Crosscorrelation function}

The crosscorrelation function (Eq.~\ref{eq:autocorrelation}) can also
be used to study the correlation between star clusters and the flux
from any image. We can rewrite Eq.~\ref{eq:autocorrelation} as:
\begin{equation}  \label{eq:crosscorrelation}
1+\xi (r) = \frac{1}{\bar{f}N}\sum_{i=1}^{N} f_{i}(r),
\end{equation}
following \citet[][their Eq.~2]{zhang01}, where $f_{i}(r)$ is now the
intensity (i.e. flux per pixel) in an aperture with radius $r$ centred on cluster $i$,
and $\bar{f}$ is the mean intensity over the whole image. For both the calculation of $f_{i}(r)$ and $\bar{f}$ the areas not covered by $U$-band imaging are masked out. To study the
crosscorrelation between star clusters and star formation, we used the
20~cm radio continuum (RC) map \citep{patrikeev06}, and the \hst/\acs\ H$\alpha$ mosaic image of M51 \citep{mutchler05}, which we continuum subtracted using a combination of the \acs\ \vband\ and \iband\ images. The 20~cm RC map was also used
by \citet{schuster07} to derive the radial SFR distribution (shown in
Fig.~\ref{fig:radial distribution}). The SFRs from the apertures of
\citet{kennicutt07} (Fig.~\ref{fig:distributions}) are not suitable to
be directly used in Eq.~\ref{eq:crosscorrelation} due to their
incomplete spatial sampling. 

The 20~cm RC is tracing star formation
through the synchrotron radiation coming from cosmic rays, emitted by
supernovae explosions of massive stars. The exact relation between SFR
and RC emission, however, is not well defined. It depends on the
scaling between RC and far-infrared dust continuum emission, of which
the general form and radial dependencies are still under debate (see
e.g. \citealt{helou85, marsh95, niklas97a, niklas97b, murgia05,
  murphy06, murphy08}). For our current study, however, we are not
interested in the exact scaling between RC or H$\alpha$ flux and SFR. We simply use
the RC and H$\alpha$ flux to study how star clusters and two spatially complete tracers
of star formation correlate for differently aged cluster samples.

\begin{figure}
\centering
 \includegraphics[width=85mm]{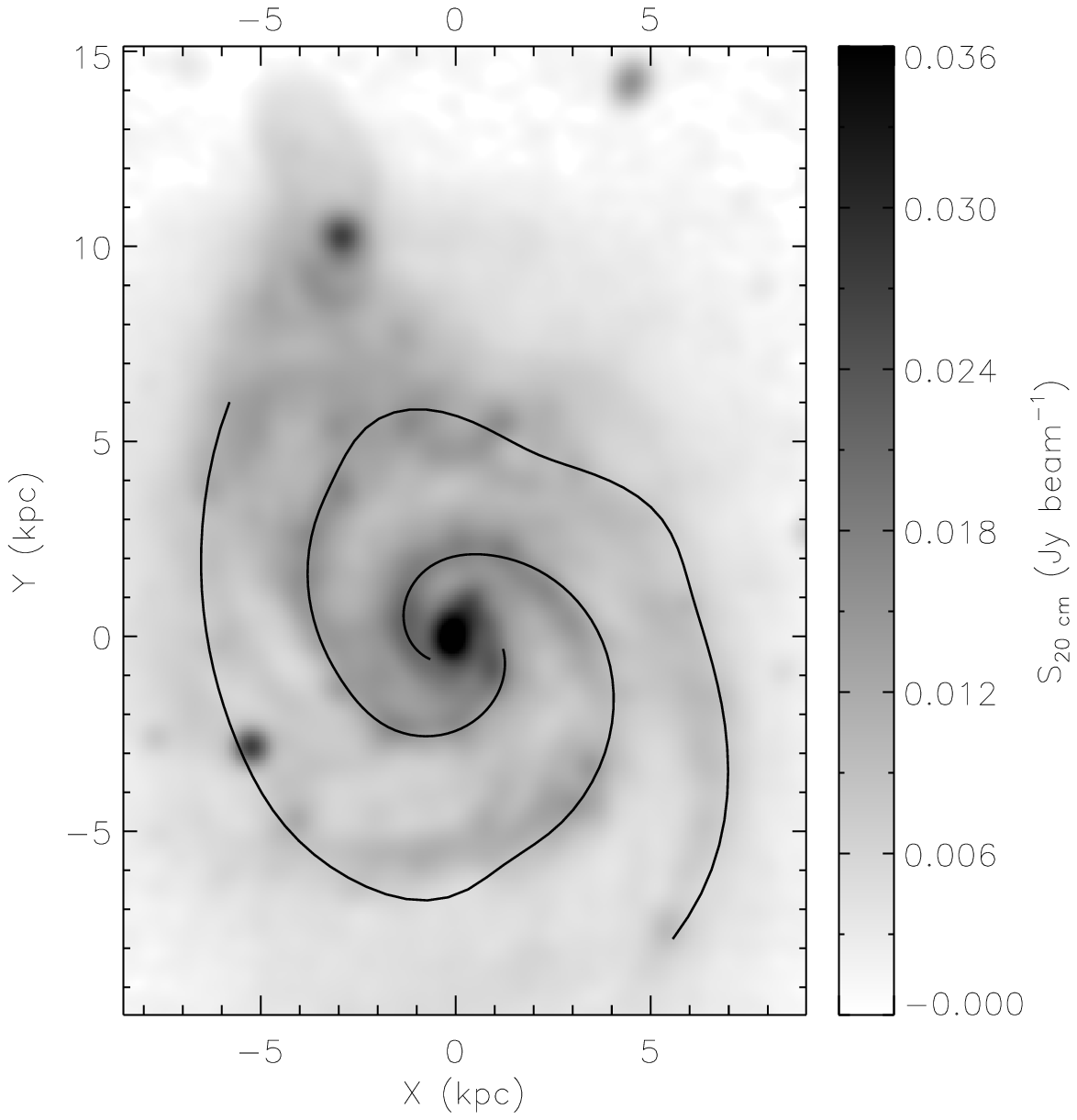} 
\caption{Radio continuum (20~cm) image of M~51, with the spiral arms of
Fig.~\ref{fig:spiralarms} overplotted for reference.}
\label{fig:radiomap}
\end{figure}

Before we applied the crosscorrelation function
(Eq.~\ref{eq:crosscorrelation}) to the 20~cm RC image (shown in
Fig.~\ref{fig:radiomap}), the image was cropped to match the
\acs\ mosaic frame and resampled to a pixel scale of $1\arcsec$ (the
original image had a pixel scale of $2\arcsec$ and a beam size of
$15\arcsec$). {Due to computational issues the H$\alpha$ image, with a resolution of $0.05\arcsec$ was binned by a factor of 10 to a pixel scale of $0.5\arcsec$.}
For the uncertainty in $1+\xi(r)$ we again followed
\citet{zhang01}, adopting only the statistical uncertainty in the
number of clusters per sample ($N$), leading to:
\begin{equation}
\delta(r) = N^{-1/2}(1+\xi(r)). 
\end{equation}

\begin{figure}
\centering
 \includegraphics[width=85mm]{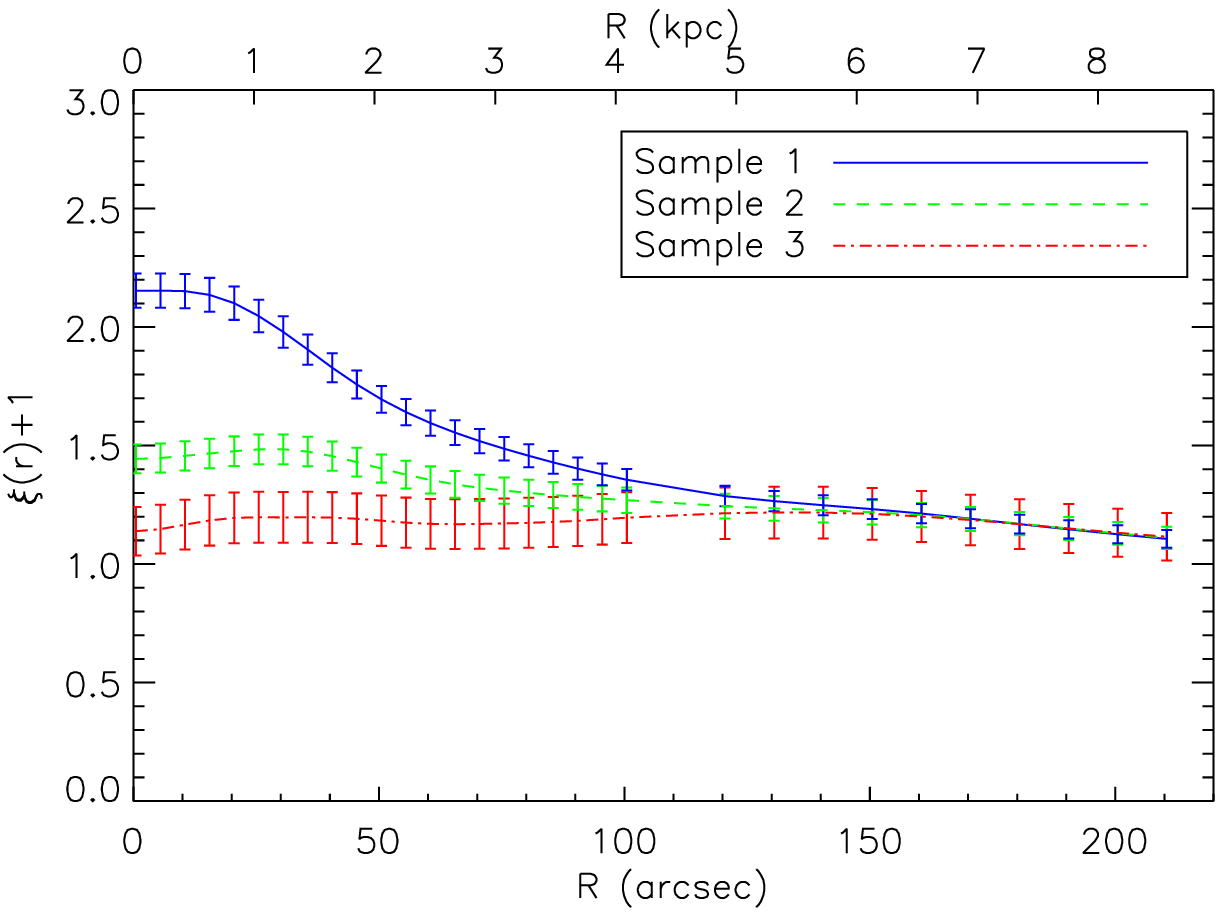} 
\caption{Crosscorrelation function between the three cluster samples
  and the 20~cm radio continuum image.}
\label{fig:crosscorrelation_radio}
\end{figure}

In Fig.~\ref{fig:crosscorrelation_radio} we show the resulting
crosscorrelation functions between the three cluster samples and the RC
flux. The crosscorrelation functions are flat on scales $< 10\arcsec$,
which is very likely a result of the beam size of the radio data. It
is evident that the youngest cluster sample is most correlated
(i.e. largest $\xi(r)$) with the RC emission on all scales smaller
than $\sim100\arcsec$ (i.e. $\sim 4$~kpc), and that the correlation
drops with the age of the cluster population. A random distribution of
clusters would have a constant $\xi(r)+1 = 1$. In other words, most of
the radio flux is concentrated around the youngest clusters, as
expected. 

The crosscorrelation function between the 20~cm RC emission and the youngest cluster sample drops to half its peak value at $R\approx 2$~kpc. This shows that the RC emission is tracing star formation in a very diffuse way, consistent with the long path length of the cosmic rays, before they stop emitting RC emission.

\begin{figure}
\centering
 \includegraphics[width=85mm]{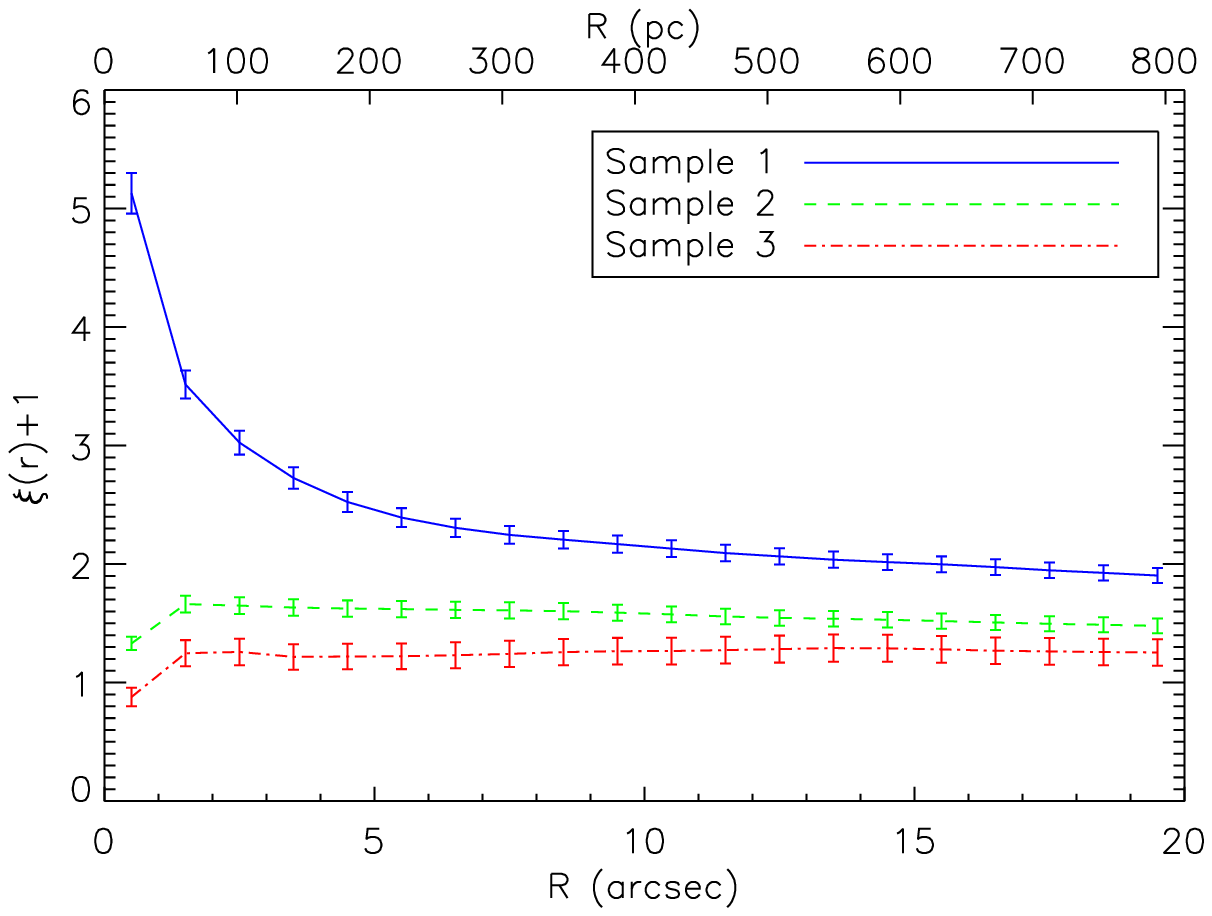} 
\caption{Crosscorrelation function between the three cluster samples and the (continuum subtracted) H$\alpha$ image. Note the difference in X- and Y-scaling compared to Figs.~\ref{fig:crosscorrelation_radio} \&~\ref{fig:crosscorrelation_2mass}. On a larger scale, the three curves would converge to $\xi+1 = 1.1$ at $R\approx120\arcsec \approx 5$~kpc, similarly to Fig.~\ref{fig:crosscorrelation_radio}.}
\label{fig:crosscorrelation_halpha}
\end{figure}

In Fig.~\ref{fig:crosscorrelation_halpha} we show the crosscorrelation functions between the three cluster samples and the (continuum subtracted) H$\alpha$ flux. The youngest clusters are strongly correlated to the H$\alpha$ flux. The correlation peaks at the resolution limit of $R = 0.5\arcsec \approx 20$~pc and drops to half the peak value at $R \approx 100$~pc. This radius is consistent with the typical expansion speed ($\sim$10~$\kms$) of the ionized bubbles of hot stars \citep[e.g.][p. 369]{lamers99}. At a radius larger than the typical travel distance of an H$\alpha$ bubble in 10~Myr (i.e. $\sim$100~pc) the correlation is expected to level off. For sample 2 clusters the correlation with H$\alpha$ is considerably less and the sample 3 clusters shows a crosscorrelation function with H$\alpha$ consistent with a random distribution (i.e. $\xi(r)+1 = 1$, although a slight offset from 1 is expected due the large scale structure or M~51, see below).

\begin{figure}
\centering
 \includegraphics[width=85mm]{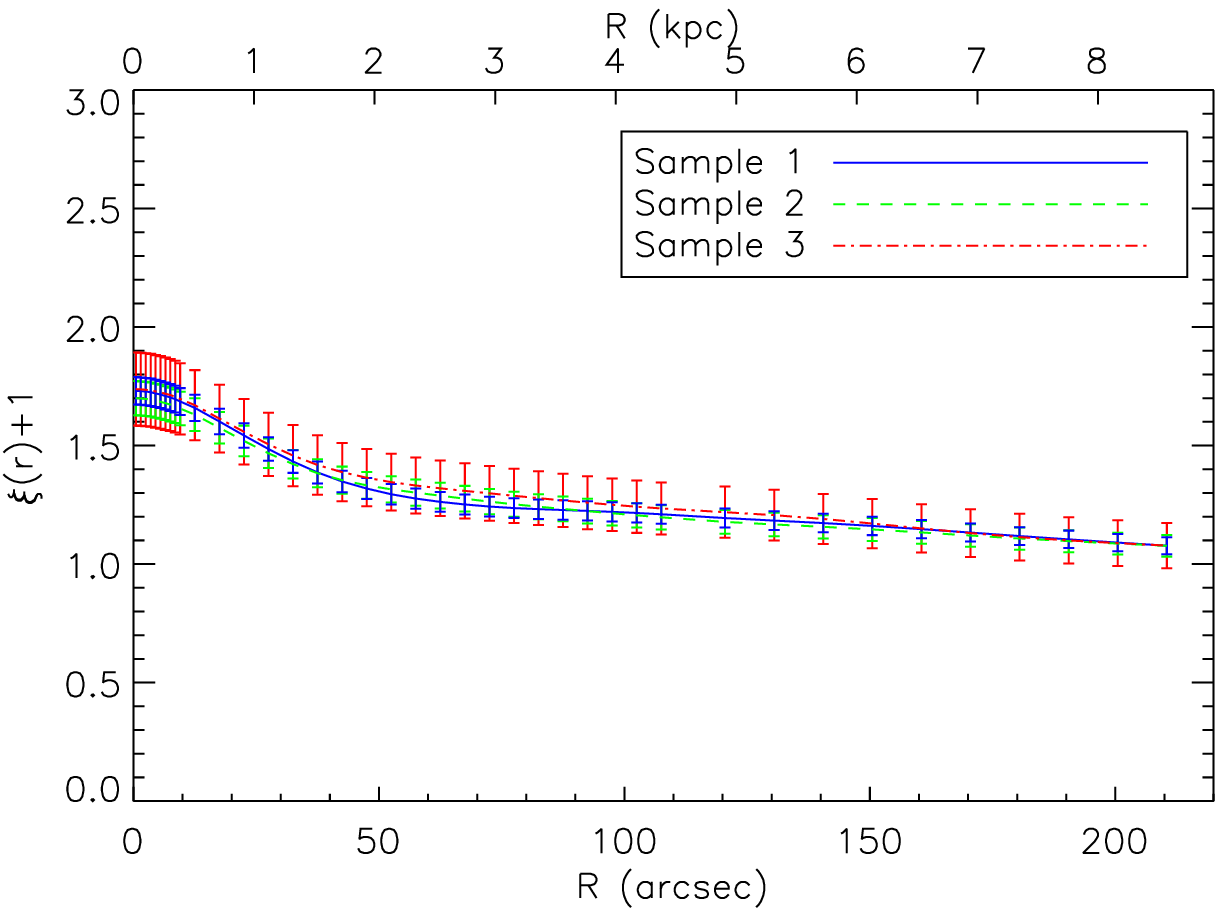} 
\caption{Crosscorrelation function between the three cluster samples and the \emph{2MASS} $H$-band image of M~51.}
\label{fig:crosscorrelation_2mass}
\end{figure}

As a consistency check, we also derived the crosscorrelation functions between the star clusters and the flux from the \emph{2MASS} $H$-band image, which we used to define the spiral arms (Fig.~\ref{fig:spiralarms}). Since this near-infrared data is mostly tracing older stellar populations, no enhanced correlation with young star clusters is expected. In Fig.~\ref{fig:crosscorrelation_2mass} the crosscorrelation functions are indeed shown to be indistinguishable. All three clusters samples, however, show an increased correlation towards smaller radii. This is likely an effect of the large scale structure of M~51. Both the cluster and $H$-band flux density are higher in the disc and spiral arms of M~51, while the observations (and thus the normalisation factor $\bar{f}$ in Eq.~\ref{eq:crosscorrelation}) cover a larger area. This leads to some degree of correlation towards smaller scales. The fact that this correlation is similar for all cluster ages confirms that the $H$-band flux is a good tracer of stellar density and not of current star formation.
  
\section{Summary and conclusions}
  \label{sec:Summary and conclusions}

We have studied the spatial correlations between star clusters and star
formation across the spiral galaxy M~51. We combined published data of the resolved star
formation law in M~51 (SFR and gas surface densities,
\citealt{kennicutt07}) with the star cluster data of
\citet{scheepmaker07}. We combined the star cluster data with new
\hst/\wfpc\ \uband\ data of 1580 clusters in order to estimate ages, masses and extinctions of these clusters using \emph{SSP} models. We used radial and azimuthal distributions and cross-correlation functions to study the spatial correlation between star formation, star clusters and spiral structure. We used
auto-correlation functions to study how the clustering of star
clusters depends on age and spatial scale. The main conclusions of this
work can be summarized as follows:

\begin{enumerate}
\item We find that age determinations of young ($\lesssim100$~Myr) star clusters based on broad-band UBVI photometry can depend strongly on the adopted stellar isochrones used in the SSP models. Using models based on the Padova isochrones, we generally find older, and therefore more massive, clusters compared to the same models based on the Geneva isochrones (Fig.~\ref{fig:age_vs_age}).

\item The radially averaged distributions of star formation and gas
  surface density in M~51 peak in the centre and subsequently at
  $\sim2.5$ and $\sim5$~kpc. These distributions are matched by the
  number surface density distribution of our youngest star
  cluster sample (ages $<10$ Myr, Fig.~\ref{fig:radial distribution}),
  while for the older clusters the distributions are more flat. 
  We note that the location of the peaks coincide approximately with
  the locations where the spiral arms meet the 4:1 resonance and the
  co-rotation radius.
  
  \item The distributions of azimuthal distances from the spiral arms show that most of the young \emph{and old} star clusters are located at the centre of the spiral arms, with the youngest star clusters showing the most smooth distribution (Fig.~\ref{fig:azimuthal}). In terms of kinematic age and for $R_{\mathrm{G}} < 4$~kpc, the azimuthal distribution of the oldest clusters peaks at the largest separation downstream of the spiral arms (10--15~Myr), while the majority of the younger clusters are within 0--5~Myr downstream of the arms. This indicates that the majority of the clusters formed $\sim5$--20~Myr before their parental gas cloud would have reached the centre of the spiral arm.
  
\item We find that the amount of clustering of star clusters decreases
  both with spatial scale and age (Fig.~\ref{fig:autocorrelation}),
  consistent with hierarchical star formation. The slope of the autocorrelation functions of the youngest clusters ($\log(\mathrm{age}) < 7.5$) is consistent with a projected fractal dimension of 1.6. We find a lower fractal dimension of 1.2 for the clusters with $\log(\mathrm{age}) < 7.0$, if we use a higher mass limit and thus, for this smaller mass range, a more complete sample. On spatial scales
  smaller than 400~pc, the amount of clustering of the \emph{oldest}
  cluster samples is very uncertain due to uncertainties in the assumed SSP model parameters.
 
 \item The locations of the youngest star clusters are highly correlated with the flux from 20~cm radio-continuum emission and continuum subtracted H$\alpha$ emission, but not with near-infrared $H$-band emission. The correlation with the radio emission is more diffuse compared to the correlation with H$\alpha$, and both correlations decreases with the age of the cluster population. 
 \end{enumerate}

In a follow-up paper \citep{scheepmaker09b} we will focus on the \emph{quantitative} comparison between star and cluster formation, discussing the fraction of total star formation taking place in optically visible star clusters.

\begin{acknowledgements}
We thank Rainer Beck, Arnold Rots and Karl Schuster for help with obtaining the radio data.
Christian Struve and Alicia Berciano Alba are thanked for some helpful discussions. We thank the anonymous referee for helpful comments that improved this paper. This research has made use of observations
made with the NASA/ESA \emph{Hubble Space Telescope}, obtained from
the data archive at the Space Telescope Institute. STScI is operated
by the association of Universities for Research in Astronomy, Inc.,
under the NASA contract NAS 5-26555. This research has also made use
of the NASA/IPAC Extragalactic Database (NED) which is operated by the
Jet Propulsion Laboratory, California Institute of Technology, under
contract with the National Aeronautics and Space Administration.
\end{acknowledgements}

\bibliographystyle{aa} 
\bibliography{11068}

\begin{thebibliography}{84}
\expandafter\ifx\csname natexlab\endcsname\relax\def\natexlab#1{#1}\fi

\bibitem[{{Anders} {et~al.}(2004){Anders}, {Bissantz}, {Fritze-v.~Alvensleben},
  \& {de Grijs}}]{anders04b}
{Anders}, P., {Bissantz}, N., {Fritze-v.~Alvensleben}, U., \& {de Grijs}, R.
  2004, \mnras, 347, 196

\bibitem[{Anders \& Fritze-v. Alvensleven(2003)}]{anders03}
Anders, P. \& Fritze-v. Alvensleven, U. 2003, A\&A, 401, 1063

\bibitem[{{Anderson} {et~al.}(1983){Anderson}, {Hodge}, \&
  {Kennicutt}}]{anderson83}
{Anderson}, S., {Hodge}, P., \& {Kennicutt}, Jr., R.~C. 1983, \apj, 265, 132

\bibitem[{{Bash}(1979)}]{bash79}
{Bash}, F.~N. 1979, \apj, 233, 524

\bibitem[{{Bastian} {et~al.}(2007){Bastian}, {Ercolano}, {Gieles},
  {Rosolowsky}, {Scheepmaker}, {Gutermuth}, \& {Efremov}}]{bastian07}
{Bastian}, N., {Ercolano}, B., {Gieles}, M., {et~al.} 2007, \mnras, 379, 1302

\bibitem[{Bastian {et~al.}(2005{\natexlab{a}})Bastian, Gieles, Efremov, \&
  Lamers}]{bastian05b}
Bastian, N., Gieles, M., Efremov, Y.~N., \& Lamers, H. J. G. L.~M.
  2005{\natexlab{a}}, A\&A, 443, 79

\bibitem[{Bastian {et~al.}(2005{\natexlab{b}})Bastian, Gieles, Lamers,
  Scheepmaker, \& de~Grijs}]{bastian05a}
Bastian, N., Gieles, M., Lamers, H. J. G. L.~M., Scheepmaker, R.~A., \&
  de~Grijs, R. 2005{\natexlab{b}}, A\&A, 431, 905

\bibitem[{{Bastian} \& {Goodwin}(2006)}]{bastian06b}
{Bastian}, N. \& {Goodwin}, S.~P. 2006, \mnras, 369, L9

\bibitem[{{Bate} {et~al.}(1998){Bate}, {Clarke}, \& {McCaughrean}}]{bate98}
{Bate}, M.~R., {Clarke}, C.~J., \& {McCaughrean}, M.~J. 1998, \mnras, 297, 1163

\bibitem[{{Beech}(1992)}]{beech92}
{Beech}, M. 1992, \apss, 192, 103

\bibitem[{{Bertelli} {et~al.}(1994){Bertelli}, {Bressan}, {Chiosi}, {Fagotto},
  \& {Nasi}}]{bertelli94}
{Bertelli}, G., {Bressan}, A., {Chiosi}, C., {Fagotto}, F., \& {Nasi}, E. 1994,
  \aaps, 106, 275

\bibitem[{Bertin \& Arnouts(1996)}]{bertin96}
Bertin, E. \& Arnouts, S. 1996, A\&AS, 117, 393

\bibitem[{{Bik} {et~al.}(2003){Bik}, {Lamers}, {Bastian}, {Panagia}, \&
  {Romaniello}}]{bik03}
{Bik}, A., {Lamers}, H.~J.~G.~L.~M., {Bastian}, N., {Panagia}, N., \&
  {Romaniello}, M. 2003, \aap, 397, 473

\bibitem[{{Boeshaar} \& {Hodge}(1977)}]{boeshaar77}
{Boeshaar}, G.~O. \& {Hodge}, P.~W. 1977, \apj, 213, 361

\bibitem[{{Calzetti} {et~al.}(2005){Calzetti}, {Kennicutt}, {Bianchi},
  {Thilker}, {Dale}, {Engelbracht}, {Leitherer}, {Meyer}, {Sosey}, {Mutchler},
  {Regan}, {Thornley}, {Armus}, {Bendo}, {Boissier}, {Boselli}, {Draine},
  {Gordon}, {Helou}, {Hollenbach}, {Kewley}, {Madore}, {Martin}, {Murphy},
  {Rieke}, {Rieke}, {Roussel}, {Sheth}, {Smith}, {Walter}, {White}, {Yi},
  {Scoville}, {Polletta}, \& {Lindler}}]{calzetti05}
{Calzetti}, D., {Kennicutt}, Jr., R.~C., {Bianchi}, L., {et~al.} 2005, \apj,
  633, 871

\bibitem[{{Cardelli} {et~al.}(1989){Cardelli}, {Clayton}, \&
  {Mathis}}]{cardelli89}
{Cardelli}, J.~A., {Clayton}, G.~C., \& {Mathis}, J.~S. 1989, \apj, 345, 245

\bibitem[{{Charbonnel} {et~al.}(1993){Charbonnel}, {Meynet}, {Maeder},
  {Schaller}, \& {Schaerer}}]{charbonnel93}
{Charbonnel}, C., {Meynet}, G., {Maeder}, A., {Schaller}, G., \& {Schaerer}, D.
  1993, \aaps, 101, 415

\bibitem[{{Condon}(1992)}]{condon92}
{Condon}, J.~J. 1992, \araa, 30, 575

\bibitem[{{Contopoulos} \& {Grosb\o l}(1986)}]{contopoulos86}
{Contopoulos}, G. \& {Grosb\o l}, P. 1986, \aap, 155, 11

\bibitem[{{Contopoulos} \& {Grosb\o l}(1988)}]{contopoulos88}
{Contopoulos}, G. \& {Grosb\o l}, P. 1988, \aap, 197, 83

\bibitem[{{de Grijs} \& {Anders}(2006)}]{degrijs06}
{de Grijs}, R. \& {Anders}, P. 2006, \mnras, 366, 295

\bibitem[{{de Grijs} {et~al.}(2005){de Grijs}, {Anders}, {Lamers}, {Bastian},
  {Fritze-v.~Alvensleben}, {Parmentier}, {Sharina}, \& {Yi}}]{degrijs05}
{de Grijs}, R., {Anders}, P., {Lamers}, H.~J.~G.~L.~M., {et~al.} 2005, \mnras,
  359, 874

\bibitem[{{de Grijs} {et~al.}(2003){de Grijs}, {Lee}, {Mora Herrera},
  {Fritze-v.~Alvensleben}, \& {Anders}}]{degrijs03d}
{de Grijs}, R., {Lee}, J.~T., {Mora Herrera}, M.~C., {Fritze-v.~Alvensleben},
  U., \& {Anders}, P. 2003, New Astronomy, 8, 155

\bibitem[{{Diaz} {et~al.}(1991){Diaz}, {Terlevich}, {Vilchez}, {Pagel}, \&
  {Edmunds}}]{diaz91}
{Diaz}, A.~I., {Terlevich}, E., {Vilchez}, J.~M., {Pagel}, B.~E.~J., \&
  {Edmunds}, M.~G. 1991, \mnras, 253, 245

\bibitem[{{Dolphin}(2000)}]{dolphin00b}
{Dolphin}, A.~E. 2000, \pasp, 112, 1397

\bibitem[{{Efremov}(1995)}]{efremov95}
{Efremov}, Y.~N. 1995, \aj, 110, 2757

\bibitem[{{Efremov} \& {Elmegreen}(1998)}]{efremov98}
{Efremov}, Y.~N. \& {Elmegreen}, B.~G. 1998, \mnras, 299, 588

\bibitem[{Elmegreen \& Efremov(1996)}]{elmegreen96b}
Elmegreen, B.~G. \& Efremov, Y.~N. 1996, ApJ, 466, 802

\bibitem[{{Elmegreen} \& {Elmegreen}(2001)}]{elmegreen01b}
{Elmegreen}, B.~G. \& {Elmegreen}, D.~M. 2001, \aj, 121, 1507

\bibitem[{{Elmegreen} {et~al.}(2006){Elmegreen}, {Elmegreen}, {Chandar},
  {Whitmore}, \& {Regan}}]{elmegreen06}
{Elmegreen}, B.~G., {Elmegreen}, D.~M., {Chandar}, R., {Whitmore}, B., \&
  {Regan}, M. 2006, \apj, 644, 879

\bibitem[{{Elmegreen} {et~al.}(1989){Elmegreen}, {Elmegreen}, \&
  {Seiden}}]{elmegreen89b}
{Elmegreen}, B.~G., {Elmegreen}, D.~M., \& {Seiden}, P.~E. 1989, \apj, 343, 602

\bibitem[{Elmegreen \& Falgarone(1996)}]{elmegreen96a}
Elmegreen, B.~G. \& Falgarone, E. 1996, ApJ, 471, 816

\bibitem[{Elson {et~al.}(1987)Elson, Fall, \& Freeman}]{elson87}
Elson, R. A.~W., Fall, S.~M., \& Freeman, K.~C. 1987, ApJ, 323, 54

\bibitem[{{Fall} {et~al.}(2005){Fall}, {Chandar}, \& {Whitmore}}]{fall05}
{Fall}, S.~M., {Chandar}, R., \& {Whitmore}, B.~C. 2005, \apjl, 631, L133

\bibitem[{Feldmeier {et~al.}(1997)Feldmeier, Ciardullo, \&
  Jacoby}]{feldmeier97}
Feldmeier, J.~J., Ciardullo, R., \& Jacoby, G.~H. 1997, ApJ, 479, 231

\bibitem[{Garc\'ia-Burillo {et~al.}(1993)Garc\'ia-Burillo, Combes, \&
  Gerin}]{garcia-burillo93}
Garc\'ia-Burillo, S., Combes, F., \& Gerin, M. 1993, A\&A, 274, 148

\bibitem[{{Gieles} \& {Bastian}(2008)}]{gieles08}
{Gieles}, M. \& {Bastian}, N. 2008, \aap, 482, 165

\bibitem[{Gieles {et~al.}(2005)Gieles, Bastian, Lamers, \& Mout}]{gieles05}
Gieles, M., Bastian, N., Lamers, H. J. G. L.~M., \& Mout, J.~N. 2005, A\&A,
  441, 949

\bibitem[{{Girardi} {et~al.}(2000){Girardi}, {Bressan}, {Bertelli}, \&
  {Chiosi}}]{girardi00}
{Girardi}, L., {Bressan}, A., {Bertelli}, G., \& {Chiosi}, C. 2000, \aaps, 141,
  371

\bibitem[{{Gomez} {et~al.}(1993){Gomez}, {Hartmann}, {Kenyon}, \&
  {Hewett}}]{gomez93}
{Gomez}, M., {Hartmann}, L., {Kenyon}, S.~J., \& {Hewett}, R. 1993, \aj, 105,
  1927

\bibitem[{{Goodwin} \& {Bastian}(2006)}]{goodwin06}
{Goodwin}, S.~P. \& {Bastian}, N. 2006, \mnras, 373, 752

\bibitem[{{Helou} {et~al.}(1985){Helou}, {Soifer}, \&
  {Rowan-Robinson}}]{helou85}
{Helou}, G., {Soifer}, B.~T., \& {Rowan-Robinson}, M. 1985, \apjl, 298, L7

\bibitem[{{Hill} {et~al.}(1997){Hill}, {Waller}, {Cornett}, {Bohlin}, {Cheng},
  {Neff}, {O'Connell}, {Roberts}, {Smith}, {Hintzen}, {Smith}, \&
  {Stecher}}]{hill97}
{Hill}, J.~K., {Waller}, W.~H., {Cornett}, R.~H., {et~al.} 1997, \apj, 477, 673

\bibitem[{{Holtzman} {et~al.}(1992){Holtzman}, {Faber}, {Shaya}, {Lauer},
  {Groth}, {Hunter}, {Baum}, {Ewald}, {Hester}, {Light}, {Lynds}, {O'Neil}, \&
  {Westphal}}]{holtzman92}
{Holtzman}, J.~A., {Faber}, S.~M., {Shaya}, E.~J., {et~al.} 1992, \aj, 103, 691

\bibitem[{Hunter {et~al.}(2003)Hunter, Elmegreen, Dupuy, \&
  Mortonson}]{hunter03}
Hunter, D.~A., Elmegreen, B.~C., Dupuy, T.~J., \& Mortonson, M. 2003, AJ, 126,
  1836

\bibitem[{{Hwang} \& {Lee}(2008)}]{hwang08}
{Hwang}, N. \& {Lee}, M.~G. 2008, \aj, 135, 1567

\bibitem[{{Jarrett} {et~al.}(2003){Jarrett}, {Chester}, {Cutri}, {Schneider},
  \& {Huchra}}]{jarrett03}
{Jarrett}, T.~H., {Chester}, T., {Cutri}, R., {Schneider}, S.~E., \& {Huchra},
  J.~P. 2003, \aj, 125, 525

\bibitem[{{Kennicutt}(1998{\natexlab{a}})}]{kennicutt98a}
{Kennicutt}, Jr., R.~C. 1998{\natexlab{a}}, \araa, 36, 189

\bibitem[{{Kennicutt}(1998{\natexlab{b}})}]{kennicutt98b}
{Kennicutt}, Jr., R.~C. 1998{\natexlab{b}}, \apj, 498, 541

\bibitem[{{Kennicutt} {et~al.}(2007){Kennicutt}, {Calzetti}, {Walter}, {Helou},
  {Hollenbach}, {Armus}, {Bendo}, {Dale}, {Draine}, {Engelbracht}, {Gordon},
  {Prescott}, {Regan}, {Thornley}, {Bot}, {Brinks}, {de Blok}, {de Mello},
  {Meyer}, {Moustakas}, {Murphy}, {Sheth}, \& {Smith}}]{kennicutt07}
{Kennicutt}, Jr., R.~C., {Calzetti}, D., {Walter}, F., {et~al.} 2007, \apj,
  671, 333

\bibitem[{{Kroupa}(2001)}]{kroupa01a}
{Kroupa}, P. 2001, \mnras, 322, 231

\bibitem[{Lada \& Lada(2003)}]{lada03}
Lada, C.~J. \& Lada, E.~A. 2003, ARA\&A, 41, 57

\bibitem[{{Lamers} \& {Cassinelli}(1999)}]{lamers99}
{Lamers}, H.~J.~G.~L.~M. \& {Cassinelli}, J.~P. 1999, {Introduction to Stellar
  Winds} (Cambridge University Press)

\bibitem[{Larsen(1999)}]{larsen99}
Larsen, S.~S. 1999, A\&A Suppl. Ser., 139, 393

\bibitem[{Larsen(2004)}]{larsen04b}
Larsen, S.~S. 2004, A\&A, 416, 537

\bibitem[{{Larsen} \& {Richtler}(1999)}]{larsen99b}
{Larsen}, S.~S. \& {Richtler}, T. 1999, \aap, 345, 59

\bibitem[{{Larsen} \& {Richtler}(2000)}]{larsen00a}
{Larsen}, S.~S. \& {Richtler}, T. 2000, \aap, 354, 836

\bibitem[{{Larson}(1995)}]{larson95}
{Larson}, R.~B. 1995, \mnras, 272, 213

\bibitem[{Mandelbrot(1983)}]{mandelbrot83}
Mandelbrot, B.~B. 1983, The fractal geometry of nature (Freeman, San Francisco)

\bibitem[{{Marsh} \& {Helou}(1995)}]{marsh95}
{Marsh}, K.~A. \& {Helou}, G. 1995, \apj, 445, 599

\bibitem[{{Meurer} {et~al.}(1995){Meurer}, {Heckman}, {Leitherer}, {Kinney},
  {Robert}, \& {Garnett}}]{meurer95}
{Meurer}, G.~R., {Heckman}, T.~M., {Leitherer}, C., {et~al.} 1995, \aj, 110,
  2665

\bibitem[{{Mihos} \& {Hernquist}(1994)}]{mihos94}
{Mihos}, J.~C. \& {Hernquist}, L. 1994, \apj, 437, 611

\bibitem[{{Murgia} {et~al.}(2005){Murgia}, {Helfer}, {Ekers}, {Blitz},
  {Moscadelli}, {Wong}, \& {Paladino}}]{murgia05}
{Murgia}, M., {Helfer}, T.~T., {Ekers}, R., {et~al.} 2005, \aap, 437, 389

\bibitem[{{Murphy} {et~al.}(2006){Murphy}, {Helou}, {Braun}, {Kenney}, {Armus},
  {Calzetti}, {Draine}, {Kennicutt}, {Roussel}, {Walter}, {Bendo}, {Buckalew},
  {Dale}, {Engelbracht}, {Smith}, \& {Thornley}}]{murphy06}
{Murphy}, E.~J., {Helou}, G., {Braun}, R., {et~al.} 2006, \apjl, 651, L111

\bibitem[{{Murphy} {et~al.}(2008){Murphy}, {Helou}, {Kenney}, {Armus}, \&
  {Braun}}]{murphy08}
{Murphy}, E.~J., {Helou}, G., {Kenney}, J.~D.~P., {Armus}, L., \& {Braun}, R.
  2008, ArXiv e-prints, 0802.2279

\bibitem[{Mutchler {et~al.}(2005)Mutchler, Beckwith, Bond, Christian, Frattare,
  Hamilton, Hamilton, Levay, Noll, \& Royle}]{mutchler05}
Mutchler, M., Beckwith, S. V.~W., Bond, H.~E., {et~al.} 2005, BAAS, 37

\bibitem[{{Niklas}(1997)}]{niklas97b}
{Niklas}, S. 1997, \aap, 322, 29

\bibitem[{{Niklas} \& {Beck}(1997)}]{niklas97a}
{Niklas}, S. \& {Beck}, R. 1997, \aap, 320, 54

\bibitem[{{Patrikeev} {et~al.}(2006){Patrikeev}, {Fletcher}, {Stepanov},
  {Beck}, {Berkhuijsen}, {Frick}, \& {Horellou}}]{patrikeev06}
{Patrikeev}, I., {Fletcher}, A., {Stepanov}, R., {et~al.} 2006, \aap, 458, 441

\bibitem[{{Peebles}(1980)}]{peebles80}
{Peebles}, P.~J.~E. 1980, {The large-scale structure of the universe} (Research
  supported by the National Science Foundation.~Princeton, N.J., Princeton
  University Press, 1980.~435 p.)

\bibitem[{{Roberts}(1969)}]{roberts69}
{Roberts}, W.~W. 1969, \apj, 158, 123

\bibitem[{{Schaerer} {et~al.}(1993){Schaerer}, {Meynet}, {Maeder}, \&
  {Schaller}}]{schaerer93}
{Schaerer}, D., {Meynet}, G., {Maeder}, A., \& {Schaller}, G. 1993, \aaps, 98,
  523

\bibitem[{{Scheepmaker} {et~al.}(2007){Scheepmaker}, {Haas}, {Gieles},
  {Bastian}, {Larsen}, \& {Lamers}}]{scheepmaker07}
{Scheepmaker}, R.~A., {Haas}, M.~R., {Gieles}, M., {et~al.} 2007, \aap, 469,
  925 (S07)

\bibitem[{{Scheepmaker} {et~al.}(2009){Scheepmaker}, {Lamers}, {Larsen}, \&
  {Anders}}]{scheepmaker09b}
{Scheepmaker}, R.~A., {Lamers}, H.~J.~G.~L.~M., {Larsen}, S.~S., \& {Anders},
  P. 2009, \aap\ (in prep)

\bibitem[{Schlegel {et~al.}(1998)Schlegel, Finkbeiner, \& Davis}]{schlegel98}
Schlegel, D.~J., Finkbeiner, D.~P., \& Davis, M. 1998, ApJ, 500, 525

\bibitem[{Schulz {et~al.}(2002)Schulz, Fritze-v. Alvensleven, M\"{o}ller, \&
  Fricke}]{schulz02}
Schulz, J., Fritze-v. Alvensleven, U., M\"{o}ller, C.~S., \& Fricke, K.~J.
  2002, A\&A, 392, 1

\bibitem[{{Schuster} {et~al.}(2007){Schuster}, {Kramer}, {Hitschfeld},
  {Garcia-Burillo}, \& {Mookerjea}}]{schuster07}
{Schuster}, K.~F., {Kramer}, C., {Hitschfeld}, M., {Garcia-Burillo}, S., \&
  {Mookerjea}, B. 2007, \aap, 461, 143

\bibitem[{{Tamburro} {et~al.}(2008){Tamburro}, {Rix}, {Walter}, {Brinks}, {de
  Blok}, {Kennicutt}, \& {Mac Low}}]{tamburro08}
{Tamburro}, D., {Rix}, H.~., {Walter}, F., {et~al.} 2008, ArXiv e-prints

\bibitem[{Tully(1974)}]{tully74}
Tully, R.~B. 1974, ApJS, 27, 449

\bibitem[{{Whitmore} {et~al.}(2005){Whitmore}, {Gilmore}, {Leitherer}, {Fall},
  {Chandar}, {Blair}, {Schweizer}, {Zhang}, \& {Miller}}]{whitmore05}
{Whitmore}, B.~C., {Gilmore}, D., {Leitherer}, C., {et~al.} 2005, \aj, 130,
  2104

\bibitem[{{Whitmore} {et~al.}(1999){Whitmore}, {Zhang}, {Leitherer}, {Fall},
  {Schweizer}, \& {Miller}}]{whitmore99b}
{Whitmore}, B.~C., {Zhang}, Q., {Leitherer}, C., {et~al.} 1999, \aj, 118, 1551

\bibitem[{{Yun} {et~al.}(2001){Yun}, {Reddy}, \& {Condon}}]{yun01}
{Yun}, M.~S., {Reddy}, N.~A., \& {Condon}, J.~J. 2001, \apj, 554, 803

\bibitem[{Zhang {et~al.}(2001)Zhang, Fall, \& Whitmore}]{zhang01}
Zhang, Q., Fall, S.~M., \& Whitmore, B.~C. 2001, ApJ, 561, 727

\bibitem[{Zimmer {et~al.}(2004)Zimmer, Rand, \& McGraw}]{zimmer04}
Zimmer, P., Rand, R.~J., \& McGraw, J.~T. 2004, ApJ, 607, 285

\end{thebibliography}

\end{document}